\documentstyle[aps,preprint,eqsecnum,epsfig]{revtex}
\tightenlines
\def\gsim{\stackrel{>}{\sim}}
\def\lsim{\stackrel{<}{\sim}}
\def\beq{\begin{equation}}
\def\eeq{\end{equation}}
\def\ba{\begin{array}}
\def\ea{\end{array}}
\def\bea{\begin{eqnarray}}
\def\eea{\end{eqnarray}}
\def\bt{\begin{table}[ht]}
\def\et{\end{table}}
\def\brt{\begin{ruledtabular}}
\def\ert{\end{ruledtabular}}
\def\btr{\begin{tabular}}
\def\etr{\end{tabular}}
\def\bc{\begin{center}}
\def\ec{\end{center}}
\def\bfg{\begin{figure}}
\def\efg{\end{figure}}

\def\abs#1{\left|#1\right|}

\begin{document}
\draft
\preprint{UFIFT-HET-04-6}

\date{May 4, 2004}
\title{Diurnal and Annual Modulation \\ of Cold Dark Matter Signals}
\author{Fu-Sin Ling$^{1,3}$, Pierre Sikivie$^1$, and Stuart Wick$^2$}

\address{$^1$ Department of Physics,
   University of Florida, \\
   Gainesville, FL,  32611 \\
   $^2$ NRC Postdoctoral Fellow, Code 7653, 
   Naval Research Lab, \\
   Washington, DC, 20375 \\
   $^3$ Service de Physique Th\'eorique, CP 225 \\
   Universit\'e Libre de Bruxelles, Brussels, Belgium}
 
\maketitle

\begin{abstract} 
We calculate the diurnal and annual modulation of the signals in axion 
and WIMP dark matter detectors on Earth caused by a cold flow of dark 
matter in the Solar neighborhood.  The effects of the Sun's and the 
Earth's gravity, and of the orbital and rotational motions of the Earth 
are included.  A cold flow on Earth produces a peak in the spectrum of 
microwave photons in cavity detectors of dark matter axions, and a plateau 
in the nuclear recoil energy spectrum in WIMP detectors. Formulas are given 
for the positions and heights of these peaks and plateaux as a function 
of time in the course of day and year, including all corrections down to 
the 0.1\% level of precision.  The results can be applied to an arbitrary 
dark matter velocity distribution $f(\vec{v})$ by integrating the one-flow 
results over velocities.  We apply them to the set of flows predicted 
by the caustic ring model of the Galactic halo.  The caustic ring model 
predicts the dark matter flux on Earth to be largest in December $\pm$ one
month.  Nonetheless, because of the role of energy thresholds, the model
is consistent with the annual modulation results published by the DAMA
collaboration provided the WIMP mass is larger than approximately 100
GeV.    
\end{abstract}


\section{Introduction}
\label{sec:int}

The existence of cold dark matter (CDM), a species of as yet unidentified
non-baryonic particle that makes up a large fraction of the energy
density of the universe, was dramatically reaffirmed recently by the
release \cite{WMAP} of the first-year data from the Wilkinson Microwave
Anisotropy Probe (WMAP).  WMAP confirmed the earlier discovery
\cite{accel} of the accelerated expansion of the universe, and determined
all cosmological parameters with unprecedented precision.  The universe is
found to be flat, with overall composition: 73\% dark energy, 23\% dark
matter, and 4.4\% ordinary baryonic matter.  [By definition, dark energy
has large negative pressure ($p/\rho \sim -1$) whereas dark matter has
negligible pressure ($p << \rho$).] The existence of dark matter had been
inferred earlier from the dynamics of galaxies and galaxy clusters
\cite{Zw,dgal}.  Dark matter is thought to be {\it collisionless}, and it
must be {\it cold} \cite{cold}.  The first attribute means that the
dynamical evolution of the dark matter is the result of purely
gravitational forces.  We believe dark matter has this property because it
reveals itself so far only through its gravitational effects and because
it forms halos around galaxies.  If dark matter were collisionful, it
would dissipate its kinetic energy after falling onto galaxies and settle
deep into the galactic gravitational wells, as ordinary matter does. The
second attribute means that the dark matter must have sufficiently low
primordial velocity dispersion to allow it to cluster easily on galactic
scales.  In contrast, neutrinos, although collisionless, have too large a
velocity dispersion to be the dark matter in galactic halos
\cite{neutrino}.

Elementary particle physics has provided two well-motivated
candidates for cold dark matter: the axion and the neutralino.  
Both were postulated for reasons wholly independent of the dark 
matter problem. The axion was postulated \cite{axion} to solve the 
strong CP problem of QCD, i.e. to explain why the strong interactions 
conserve P and CP in spite of the fact that the Standard Model as a 
whole violates those symmetries.  It has the required cosmological 
energy density to be the dark matter if its mass $m_a$ is of order 
10 $\mu$eV \cite{axcosm}.  In this mass range, it is extremely weakly
interacting.  The primordial velocity dispersion of dark matter axions 
is very small:
\beq
\delta v_a (t) \sim 3 \cdot 10^{-17}~c~
\left({10^{-5}~{\rm eV} \over m_a}\right)^{5 \over 6}~
\left(t_0 \over t\right)^{2 \over 3}~~~\ ,
\label{dva}
\eeq
where $t$ is the age of the universe, and $t_0$ the present age.
The neutralino is a dark matter candidate \cite{neutralino} suggested 
by supersymmetric extensions of the Standard Model.  Supersymmetry has 
been postulated to solve the hierarchy problem, i.e. to explain why the 
electroweak scale is small compared to other known scales such as the 
Planck mass, or the grand unification scale.  Neutralinos are examples 
of a broader class of cold dark matter candidates called WIMPs
\cite{WIMP}.  WIMP stands for 'weakly interacting massive particle'.  
The neutralino mass is of order 100 GeV.  The primordial velocity 
dispersion of WIMPs is
\beq
\delta v_W(t) \sim 10^{-11}~c~
\left({{\rm GeV} \over m_W}\right)^{1 \over 2}~
\left(t_0 \over t\right)^{2 \over 3}~~~\ ,
\label{dvW}
\eeq
where $m_W$ is the WIMP mass.  

Fortunately, both dark matter candidates can be detected on Earth, at
least in principle.  As was mentioned already, CDM falls into the
gravitational wells of galaxies and forms invisible halos around their
luminous parts.  The evolution and structure of galactic halos can be
modeled.  Although there is at present no consensus on the validity of
specific models, it is generally agreed that the density of CDM on Earth
is of order $10^{-24}$ gr/cm$^3$.  Dark matter axions on Earth can be
detected \cite{axdet,ADMX} by stimulating their conversion to microwave
photons in an electromagnetic cavity permeated by a large static magnetic
field.  The energy of the outgoing microwave photon equals that of the
incoming axion.  The spectrum of microwave photons in the cavity can be
measured very accurately.  So, if a signal is found, it becomes possible
to measure the kinetic energy spectrum of CDM on Earth with great
precision.  Note that all CDM candidates have the same velocity
distribution at any physical location because the velocities of cold
collisionless particles are the outcome of purely gravitational
interactions.  WIMP particles can be detected on Earth by measuring the
recoil energy of nuclei off of which WIMPs have scattered elastically in
the laboratory \cite{Wdet}.  Unfortunately, because the scattering angle
is not measured, one cannot deduce the kinetic energy of the incoming WIMP 
on an event by event basis.

Any signal of CDM in the laboratory will depend critically upon the
details of the local halo velocity distribution \cite{Wrev}.  The
oldest and simplest model of galactic halos is the isothermal model
\cite{iso}.  It assumes that the dark matter particles form a 
self-gravitating sphere in thermal equilibrium.  The model depends 
on only two parameters:  the halo velocity dispersion and a core 
radius.  It predicts that the local velocity distribution is Maxwellian 
in the non-rotating restframe of the Galaxy with velocity dispersion 
$\delta v_{\rm halo} \simeq 270$ km/s.  In that case the flux of CDM 
particles on Earth is largest near June 2 \cite{ann}, when the orbital
motion of the Earth is most nearly in the same direction as the motion 
of the Sun around the Galaxy, and smallest six months later.  

However, the underlying assumption of the isothermal model, that 
present galactic halos are thermalized, is not justified.  Even if 
a halo was thermalized during an early era of "violent relaxation"
\cite{vio}, the continuing infall of CDM particles into the galactic
gravitational well soon adds a non-thermal component to the halo 
\cite{ips}. An isolated galaxy, such as our own, keeps growing by 
the infall of the surrounding dark matter.  The dark matter flows 
which result from this continual infall do not thermalize over the 
age of the universe if gravity is the only force acting on the dark 
matter.  It was shown in refs. \cite{ips,stw} that the fraction of the
local
dark matter density in these discrete flows is large (of order one).
The flows form peaks in velocity space \cite{ips} and caustics 
\cite{PS1,PS2,Trem} in physical space. 

$N$-body simulations \cite{sim} correctly describe the evolution and
structure of galactic halos provided $N$ is large enough.  How large must
$N$ be?  The number of particles per galactic halo is of order $10^{84}$
in the case of axions, and $10^{68}$ in the case of WIMPs.  Because they
have negligible primordial velocity dispersion, CDM particles lie on a
3-dim. sheet in 6-dim. phase-space.  The problem of understanding the
evolution and structure of galactic halos is identical to the problem of
understanding the folding of this phase-space sheet.  The minimum number
of folds of the phase-space sheet at the Sun's location in the Galaxy
(equivalently, the minimum number of CDM flows on Earth) is of order 100
\cite{ips}.  Because phase-space is 6-dim. it takes at least $100^6 =
10^{12}$ particles to describe the phase-space structure of the Milky Way
halo down to our location in it, 8.5 kpc from the Galactic center.  This
is a strict kinematical requirement. Present simulations have only of
order $10^7$ particles and hence can at best give a crude description of
galactic halos.  Present simulations are also afflicted by two-body
relaxation \cite{2body}: the simulated particles (each representing
approximately $10^7$ solar masses) make hard two-body collisions with one
another.  (Two-body interactions between axions or WIMPs are of course
entirely negligible.)  Note that the formation of discrete CDM flows is
evident in simulations \cite{stiff} when appropriate care is taken.

A model of the structure of isolated galactic halos \cite{stw,PS1,PS2,bux}
has been developed which explicitly realizes the fact that all CDM
particles lie on a 3-dim. sheet in phase-space.  Called the 'caustic ring
halo model', it is described in some detail in Appendix B.  It predicts a
discrete set of flows at any physical location.  The number of flows is
location dependent.  Caustic surfaces separate the regions with a
differing number of flows.  At the caustics, the dark matter density is
very large.  Evidence for the caustics predicted by the model has been
found in observations of the Milky Way \cite{mw} and of other galaxies
\cite{kin}.  From fitting the model to the observations, a set of CDM
flows at the Earth's location has been determined.  The flows are listed
in Table I.  It is an update of a similar table published in ref.
\cite{bux}.  Because of our proximity to a caustic, a pair of flows
dominates the local CDM distribution.  It is given by the $n=5$ entry in
Table I.  One member of the pair, dubbed the 'big flow', has of order
three times as much density as all the other flows combined.

It is not the purpose of our paper to review galactic halo models, 
or to try to settle the associated controversies.  We described 
the main approaches to galactic halo modeling only to provide a 
context for our work, whose aim is much more modest.  We wish to 
answer the question: given a single flow of dark matter of uniform 
density $d_0$ and velocity $\vec{v}_0$ far from the Sun, what 
signals are produced in a dark matter detector on Earth as a
function of time of day and of year?  Our answer below takes 
account of the orbital and rotational motions of the Earth, and 
the gravitational fields of the Sun and the Earth.  By including
these effects, an accuracy of 0.1\% in the answer is achieved.  

The effect of solar gravity \cite{SW} is such that a single uniform flow
incident upon the Sun produces three flows downstream of the Sun, whereas
there is only one flow upstream.  See Fig. 1.  The surface separating the
one-flow region from the three-flow region is the location of a caustic,
called the 'skirt'.  It is a cone centered on the Sun, with opening angle
equal to the maximum scattering angle of the particles in the flow.  
There is another caustic, called the 'spike', on a line downstream of the
Sun, where the incoming flow gets focused by the gravity of the Sun. For
the sake of clarity, we call the initial flow, with uniform density $d_0$
and velocity $\vec{v}_0$ far from the Sun, the 'mother flow'.  The mother
flow produces near the Sun one or three position-dependent 'daughter
flows' with densities $d_i(\vec{r})$ and velocities $\vec{v}_i(\vec{r})$,
where $i=1$ or $i=1,2,3$.  The present axion and WIMP dark matter
detectors are direction-independent.  The signals produced in these
detectors depend only on the densities $d_i(t)$ and speeds $v_i(t)$ of the
daughter flows in the laboratory rest frame as functions of time of day
and of year.  For given $d_0$ and $\vec{v}_0$, we calculate $d_i(t)$ and
$v_i(t)$ with the precision advertised above.

We have been motivated by the expectation that the local CDM 
velocity distribution is dominated by a set of discrete flows, 
as in the caustic ring halo model.  Our results can be applied 
directly to such flows.  Our results can also be applied directly 
to local streams of dark matter caused by the tidal disruption of 
small satellite galaxies captured by the Milky Way \cite{stream}.  
Observations and modeling of the tidal disruption of the Sagittarius 
A dwarf galaxy by the gravitational field of the Milky Way indicates 
that the Sun is very close to that stream, and may be in it
\cite{SagA}.  Finally, we stress that our results can be applied 
to any local CDM velocity distribution $f(\vec{v}_0)$ by simply 
integrating the single mother flow results over the initial velocity 
$\vec{v}_0$.

It was already mentioned that the isothermal model predicts the flux of
dark matter on Earth to be largest near June 2, when the orbital motion
of the Earth is most nearly in the same direction as the motion of the Sun
around the Galactic center.  The reason is that, in the isothermal model,
the dark matter halo is not rotating.  The dark matter particles in the
isothermal model carry no net angular momentum.  This is unrealistic
because the angular momentum exhibited by the luminous matter is
presumably caused by gravitational forces which must have also applied a
net torque to the dark matter.  The reasonable expectation is that the
dark halo is rotating in the same direction as the visible matter.  When
halo rotation is included, it is not at all clear that the flux of CDM on
Earth is largest on June 2, and lowest six months later, as predicted by
the isothermal model.  In fact the caustic ring halo model, which includes
net angular momentum of the dark matter particles in a systematic way,
predicts that the CDM flux is largest in December $\pm$ one month.  In the
caustic ring halo model, the CDM flows of highest local density have
velocities in the direction of galactic rotation and with magnitude
($\sim$ 480 km/s) larger than the speed (220 km/s) of the visible matter
\cite{bux}.  In that case the annual modulation of the CDM flux on Earth
has phase opposite to the prediction of the isothermal model.

Motivated by this remark, J. Vergados \cite{ver}, A. Green \cite{gre},
and G. Gelmini and P. Gondolo \cite{gel} investigated the annual
modulation in the caustic ring halo model of the signal searched for by
WIMP dark matter experiments.  In these pioneering papers, it was
established that a flow of dark matter on Earth produces a plateau in the
spectrum of recoil energies $E_r$ of nuclei off of which WIMPs have
scattered
elastically.  The edge of the plateau is the maximum nuclear recoil energy
$E_{\rm max}$ for given WIMP mass $m_\chi$ and flow velocity $v_F$ through
the laboratory.  $E_{\rm max}$ is proportional to $v_F^2$.  The height of
the plateau is proportional to $v_F^{-1}$.  The total signal, obtained by
integrating the recoil energy spectrum over recoil energies, is therefore
proportional to $v_F^2 \cdot v_F^{-1}~=~v_F$, i.e. it is proportional to
the flux of WIMPs through the laboratory as expected.  So the total WIMP
signal in the caustic ring halo model is largest in December, when the CDM
flux is largest.  However, as the above-mentioned authors \cite{ver,gre,gel} 
already emphasized, actual WIMP search experiments are generally only
sensitive to a finite range of possible recoil energies, say from $E_1$ 
to $E_2$.  Depending on the relative values of $E_1$, $E_2$ and 
$E_{\rm max}$, the measured WIMP rate may be proportional to $v_F$, or 
inversely proportional to $v_F$, or something else altogether.

The DAMA WIMP dark matter experiment in Gran Sasso Laboratory observes an
annual modulation of its event rate, and interprets the effect as due to
WIMPs \cite{DAMA}.  The DAMA event rate is maximum on May 21 $\pm$ 22
days, in agreement with the modulation of the WIMP flux in the isothermal
model (maximum on June 2).  The region of parameter space claimed by DAMA
appears inconsistent with the null observations of the CDMS \cite{cdms},
Edelweiss \cite{edel} and Zeplin \cite{zep} experiments.  However, because
the latter do not use the same target nuclei as DAMA, the comparison
necessarily involves some assumptions about the WIMP couplings as well as
assumptions about the structure of the Milky Way halo.  We emphasized
already the lack of consensus on the validity of any specific halo model.  
We also mentioned that the caustic ring model predicts that the CDM flux
is largest in December $\pm$ one month.  Such a prediction may at first
sight appear inconsistent with the DAMA results but, because DAMA observes
recoil energies over a finite range only, depending on the WIMP mass, and
for the reasons mentioned at the end of the previous paragraph, there is,
in fact, no such inconsistency.  A. Green concluded \cite{gre} that the
caustic ring model, as described in \cite{bux}, and the DAMA results are
consistent if the WIMP mass is larger than 30 GeV.

The present paper is guided by the following considerations. First, there
is strong need for a detailed analysis of the annual and diurnal
modulations of the signals in the axion dark matter experiment, with a
precision commensurate with that of the experimental apparatus.  If a
signal is found in the ADMX cavity detector \cite{ADMX} presently taking
data at Lawrence Livermore National Laboratory, the CDM kinetic energy
spectrum will be measured readily with a precision of $10^{-5}$, and this
would soon be improved upon.  It is desirable therefore to calculate the
diurnal and annual modulation of the CDM density and velocity distribution
on Earth with comparable precision.  To this end we take account not only
of the orbital motion of the Earth, but also the effect of solar gravity,
the rotation of the Earth, and the effect of Earth's gravity.  We use the
previous results by two of us \cite{SW} on the effect of solar gravity.  
Second, since the axion and WIMP velocity distributions are the same, we
present the results in a way which can be readily used by the WIMP dark
matter experiments as well.  The precision and sensitivity of the WIMP
experiments is rapidly improving.  If a candidate WIMP signal is found,
the observation of the diurnal modulation and/or the effect of Solar
gravity may provide an unambiguous confirmation.  Third, we update the
predictions of the caustic ring model in refs. \cite{ver,gre,gel} because
the caustic ring model has evolved since those papers appeared.  In ref.
\cite{mw} the model was fitted to measurements of the Milky Way rotation
curve and to a triangular feature in the IRAS map of the galactic plane.  
The IRAS feature is interpreted to be the imprint of the nearest caustic
ring of dark matter on gas and dust in the galactic plane.  If so, the Sun
is very close to a cusp in the caustic ring nearest to us.  This implies
that the densities of the flows involved in that caustic (the $n=5$ flows)
are very large at our location. The fit requires an update of the list of
local CDM flows predicted by the model.  The new list of flows is given in
Table I.  The main difference with ref. \cite{bux} is that the fifth flows
($n=5$) are now much more prominent.  In fact, the revised caustic ring
halo model predicts that the local dark matter velocity distribution is
dominated by a single flow, the 'big flow', containing of order 75\% or
more of the local dark matter density.

The remainder of this paper is organized as follows.  In Section II we 
review the results of ref. \cite{SW} on the effect of the Sun's gravity
on a cold collisionless dark matter flow.  Sec.~III describes the 
diurnal and annual modulation of the signals in the axion dark matter 
cavity experiment.  Sec.~IV describes the modulation of the WIMP 
signal.  The predictions of the caustic ring halo model are compared 
with the DAMA results.  We give our concluding remarks in Sec.~V.  
Appendices describe our coordinate systems and the local CDM velocity 
distribution predicted by the caustic ring model.  

\section{Effect of the Sun's Gravity}
\label{sec:sun}

This paper analyzes the diurnal and annual modulation of the signals 
in dark matter detectors located on Earth due to discrete cold flows 
of dark matter in the Solar neighborhood.  The annual modulation is 
caused by the Earth's orbital motion and by the spatially varying 
effect of the Sun's gravity on the flow.  We summarize in this 
section relevant results from ref. \cite{SW} describing the effect 
of the Sun's gravity.

We assume the flow to be spatially and temporally uniform far 
upstream of the Sun with density $d_0$ and velocity $\vec{v}_0$.
We assume also that the velocity dispersion of the flow is 
negligibly small.  The effect of the Sun's gravity is such that 
there are three flows ($i=1,2,3$) downstream, whereas there is 
only one flow ($i=1$) upstream.  See Fig. 1.  To describe the 
flows, it is convenient to adopt cylindrical coordinates 
($z,\rho,\phi$) such that $\vec{v}_0 = v_0 \hat{z}$, with 
$0 \leq \phi < \pi$ and $-\infty < \rho < +\infty$ (instead 
of the usual $0 \leq \phi < 2\pi$ and $ 0 \leq \rho <+\infty$).
The range of $z$ is $-\infty < z < +\infty$ as usual.  The flows 
at a given location $(z, \rho, \phi)$ have different impact parameters
$b_i (z,\rho)$.  We allow the $b_i$ to have either sign as well.  For 
flow 1, the impact parameter $b_1$ has the same sign as $\rho$, whereas
the impact parameters $b_2$ and $b_3$ of the other two flows have
opposite sign from $\rho$.  See Fig. 1 . Flow 1 exists both upstream 
and downstream of the Sun, whereas flows 2 and 3 exist only downstream.
Each flow is scattered through an angle $\Theta$ where 
$-\Theta_{\rm max} \leq \Theta \leq +\Theta_{\rm max}$ and
$\Theta_{\rm max}$ is the maximum scattering angle among all 
trajectories.  Flows 2 and 3 are present downstream at angles 
$-\Theta_{\rm max} \leq \theta \leq +\Theta_{\rm max}$ and end 
on a conical caustic called the 'skirt', with opening angle
$\Theta_{\rm max}.$  In the limit of a point mass Sun, 
$\Theta_{\rm max} \rightarrow \pi$, flows 1 and 2 exist everywhere, 
and flow 3 disappears because $d_3 \rightarrow 0$ everywhere in that 
limit.  Flow 3 always goes through the Sun.  In addition to the skirt,
there is a caustic located on a line downstream of the Sun, called the 
'spike'.  It occurs because the gravitational field of the Sun focuses 
the (collisionless) flow of dark matter.

As a function of position ($z, \rho, \phi$) the impact parameters 
of flows 1 and 2 are
\begin{equation}
b_{1 \atop 2} (z, \rho) = 
{\rho \over 2} \left(1 \pm \sqrt{1 + {4a(r + z) \over \rho^2}}\right)~~~,
\label{b12}
\end{equation}
where $a = {G M_\odot \over v_0^2}$, and $r = \sqrt{\rho^2 + z^2}$.
Eq. (\ref{b12}) assumes that the mass distribution of the Sun is
spherically symmetric and that the particles comprising the flows 
in question did not go through the Sun at any point in the past.  
The latter condition is satisfied at most locations in the solar
neighborhood, but not everywhere.  Note that the quadratic equation:
\begin{equation}
b^2 - b\rho - a(r + z) = 0
\label{qu}
\end{equation}
is solved by $b_1$ and $b_2$.

The properties of the flows which have previously gone through the 
Sun can be expressed most readily in terms of the scattering angle
$\Theta(b_{\rm s})$. $b_{\rm s}$ is the impact parameter of the 
trajectory of scattering angle $\Theta$.  We take $b_{\rm s}$ to 
be always positive, the usual convention.  The relation between 
$\Theta$ and $b_{\rm s}$ depends for small values of $b_{\rm s}$ 
on the mass distribution inside the Sun.  $\Theta(b_{\rm s})$
was derived in ref. \cite{SW} for a particular solar model.  The 
relation is two to one: for every value of the scattering angle 
there are two values of the impact parameter.  Let us call them 
$b_{{\rm s}2}(\Theta)$ and $b_{{\rm s}3}(\Theta)$, with 
$b_{{\rm s}3}(\Theta) < b_{{\rm s}2}(\Theta)$; see Fig. 2.  
Sufficiently far from the Sun, i.e. for $r >> R_\odot$ where 
$R_\odot$ = $6.961 \cdot 10^{10}$ cm is the solar radius, the 
scattering angle $\Theta$ of the particles in a flow that passed 
through the Sun is approximately the polar angle 
$\theta = \tan^{-1}({\rho \over z})$ at that location. For such
flows ($i = 2,3$) 
\begin{equation} |b_i(\theta)| = b_{{\rm s}i}
(\Theta = \theta)  + 0 ({R_\odot \over r})~~~\ . 
\label{b23}
\end{equation} 
Far from the Sun, such as on Earth, the flows that passed
through the Sun, and to which Eq. (\ref{b23}) therefore applies, 
are flow 3 everywhere and flow 2 for $\theta$ near $\Theta_{\rm
max}$.  Fig. 3 shows the impact parameters $b_i(\theta)$ for a 
particular solar model and a particular value of the initial speed
$v_0$.  Note that at $\theta = 0$ flow 1 becomes flow 2 and vice versa.

We now describe the flows by giving their density and velocity 
vector as functions of position.  For all flows ($i = 1,2,3$), 
we may write the following expression for the velocity
\begin{equation}
\vec{v}_{\odot i} (\vec{r}) = \pm \hat{r} v_0 \sqrt{1 + {2a \over r} - 
({b_i \over r})^2} - \hat{\theta} v_0 {b_i \over r}~~~~~\ .
\label{vi}
\end{equation}
Eq. (\ref{vi}) follows from energy and angular momentum conservation 
and is valid everywhere outside the Sun.  It is also valid inside 
the Sun provided we replace ${2a \over r}$ by ${- 2 V(r) \over v_0^2}$, 
where $V(r)$ is the Sun's gravitational potential.  The $+(-)$ sign 
pertains to where the flow is outgoing (incoming), i.e. where 
$\dot{r} > 0$ ($\dot{r} < 0$).  For the flows that did not go through 
the Sun, Eq. (\ref{vi}) may be rewritten in other useful ways ($i = 1,2$):
\begin{eqnarray}
\vec{v}_{\odot i}(\vec{r}) &=& 
\hat{z} v_0 (1 + {a \rho \over r b_i})
- \hat{\rho} {v_0 \over r} (b_i - \rho)\nonumber\\
&=&\vec{v}_0~+~ {1 \over 2} v_0 
\left(1~\mp~\sqrt{1+Y}\right)(\hat{r} - \hat{z})~~~\ .
\label{v12}
\end{eqnarray}
The densities of the flows that did not go through the Sun are
\begin{equation}
d_{1 \atop 2}(\vec{r}) = {d_0 \over 4}
\left(\sqrt{1 + Y} + {1 \over \sqrt{1 + Y}} \pm 2 \right)
\label{d12}
\end{equation}
where
\begin{equation}
Y = {4a \over \rho^2} (r + z) = 
{2a \over r \sin^2({\theta \over 2})}~~~~\ .
\label{Y}
\end{equation}
The spike caustic is evident because $Y$ diverges as ${8ar \over \rho^2}$
when $\rho \rightarrow 0$ with $z = +r$, implying $d_{1,2}
\rightarrow \sqrt{ar \over 2} {d_0 \over \rho}$ in that limit.
The densities of the flows that did go through the Sun are ($i = 2,3$)
\begin{equation}
d_i= {d_0 \over r^2 |\sin \theta|}
\left|{b_{{\rm s}i}(\theta) \over 
{d\Theta \over db_{{\rm s}i}}(\theta)}\right|
(1 + 0 ({R_\odot \over r}))~~~\ .
\label{d23}
\end{equation}
The skirt caustic is evident because ${d\Theta \over db_{\rm s}}$
goes to zero at the maximum scattering angle $\Theta_{\rm max}$.  

\section{Axion signal modulation}
\label{sec:axion}

In this section, we discuss the diurnal and annual modulation 
of a signal in a cavity detector of dark matter axions due to 
a single cold flow of axions incident upon the solar system.  
We assume the detector is placed on Earth.  

The detector is an electromagnetic cavity permeated by a strong 
static magnetic field \cite{axdet}.  In this cavity galactic halo 
axions convert to microwave photons.  The microwave power is amplified 
and spectrum analyzed.  For every conversion, the photon energy equals 
the total (rest mass + kinetic) energy of the incoming axion:
\begin{equation}
\hbar \omega = {m_a c^2 \over \sqrt{1 - ({v \over c})^2}} =
m_a c^2 \left(1 + {1 \over 2}({v \over c})^2 + 
{3 \over 8} ({v \over c})^4 + ... \right)
\label{om}
\end{equation}
where $m_a$ is the axion mass and $v$ is the axion speed
in the laboratory frame.  The conversion is resonantly 
enhanced if the photon frequency $f = {\omega \over 2 \pi}$
falls within the cavity bandwidth.  Dark matter axions are 
expected to have a mass in the 
$10^{-5}$ eV/$c^2$ = $2 \pi (2.4179895$ GHz)${\hbar \over c^2}$
range with large uncertainties.  Galactic halo axions have 
velocities of order $10^{-3}c$.  Hence the spectrum of 
microwave photons from $a \rightarrow \gamma$ conversion 
in the cavity detector has width of order $10^{-6} \nu_a$
above $\nu_a \equiv {m_a c^2 \over 2 \pi \hbar}$.

The microwave power from $a \rightarrow \gamma$ conversion 
is proportional to the quality factor $Q$ of the cavity.
Because it must be permeated by a strong magnetic field, 
the cavity is made of normal (i.e. non-superconducting)
material.  This allows $Q \sim 10^5$.  In signal search 
mode, the cavity is tuned over a wide frequency range, in 
the hope of finding $\nu_a$.  At a given frequency, the 
detector is sensitive to dark matter axions over a 
frequency range equal to the cavity bandwidth $f/Q$.
Hence, if for example one octave in frequency (from $f$ 
to $2f$) is to be covered in one year, the amount $t$ of 
time that can be spent at each cavity frequency is of order 
100 sec.  This in turn sets the maximum frequency resolution 
of the cavity detector in search mode to be of order 10 mHz.  
When this highest frequency resolution is achieved - as is 
presently the case with the ADMX detector \cite{ADMX} at 
Lawrence Livermore National Laboratory - the energy resolution  
\begin{equation}
{\Delta E \over E} = {v \Delta v \over c^2} = {\Delta f \over f}
\label{re}
\end{equation}
is of order $10^{-11}$ and hence the velocity resolution 
\begin{equation}
\Delta v \sim 10^{-11} {c^2 \over v} \sim 10^{-8}c = 3~{\rm m/s}~~~\ .
\label{vr}
\end{equation}
Note that if a signal is discovered the detector will remain 
at the axion mass frequency for much longer times, and the 
resolution will be correspondingly increased.

Consider now the signal from a single cold flow of dark matter 
axions incident upon the solar system.  It produces a peak 
in the frequency spectrum of microwave photons from axion to 
photon conversion in the cavity detector.  The size of the 
peak is proportional to the density of the flow on Earth.
The frequency of the peak is given by Eq. (\ref{om}).  
The width of the peak is
\begin{equation}
\delta f = f {v \delta v \over c^2} 
\sim 0.3 \cdot 10^{-11} \nu_a \left({\delta v \over {\rm m/s}}\right)
\label{df}
\end{equation}
where $\delta v$ is the velocity dispersion of the flow at 
the detector location.  As the Earth orbits the Sun, both 
the density $d$ and the flow velocity $\vec{v}$ change, 
producing an annual modulation of the signal.  In addition, 
the rotation of the Earth produces a diurnal modulation of 
the signal frequency.  We will discuss first the frequency 
modulation and then the size modulation.

\subsection{Frequency modulation}

We saw in Section II, that a single flow, of uniform density $d_0$ and 
velocity $\vec{v}_0$ far from the Sun, produces one or three flows 
near the Sun depending on location, and hence one or three peaks 
in the cavity detector of dark matter axions depending on time of 
year and on the direction of $\vec{v}_0$ relative to the Earth's orbit.
Eq. (\ref{vi}) gives the flow velocities $\vec{v}_{\odot i}$ in the rest 
frame of the Sun.  The velocities in the detector rest frame are
\begin{equation}
\vec{v}_i = \vec{v}_{\odot i} - \vec{v}_{\rm orb}(t)
- \vec{v}_{\rm rot}(t) + \Delta \vec{v}_{i \oplus}
\label{ve}
\end{equation}
where $\vec{v}_{\rm orb}(t)$ is the orbital velocity of the Earth, 
$\vec{v}_{\rm rot}(t)$ is the velocity component of the laboratory 
caused by Earth's rotation, and $\Delta \vec{v}_{i \oplus}$ is the 
effect of the Earth's gravity.  We neglect here the effect of the 
gravitational fields of the other planets.  The frequency shift due 
to Jupiter is of order ${\Delta f \over f} \sim {G M_J \over c^2 R_J^2}
\sim 10^{-12}$ where $M_J$ is Jupiter's mass and $R_J$ its average  
distance from the Sun.  Such effects are certainly measurable in the 
cavity detector but they are small and complicated to describe in 
detail.  So we ignore them in this paper.  The relativistic corrections 
in Eq. (\ref{om}) are also of order $10^{-12}$, and we neglect them
as well.

\subsubsection{Effect of the Earth's gravity}

We have therefore
\begin{eqnarray}
{\Delta f_i \over \nu_a} \equiv {f_i - \nu_a \over \nu_a}
&=& {1 \over 2 c^2} (\vec{v}_{\odot i} - \vec{v}_{\rm orb} 
- \vec{v}_{\rm rot} + \Delta \vec{v}_{i \oplus})^2\nonumber\\
&=& {1 \over 2 c^2} (\vec{v}_{\odot i} - \vec{v}_{\rm orb} 
- \vec{v}_{\rm rot})^2 + {G M_\oplus \over c^2 R_\oplus}
\label{Df}
\end{eqnarray}
where $M_\oplus$ and $R_\oplus$ are the mass and radius of the 
Earth.  The last equality in Eq. (\ref{Df}) follows from energy 
conservation.  The effect of Earth's gravity is to increase
all peak frequencies by the same amount
\begin{equation}
{\Delta f_i \over \nu_a}\Bigg|_{\oplus {\rm g}} 
= {G M_\oplus \over c^2 R_\oplus} = 6.96~10^{-10}
\label{ea}
\end{equation}
independently of time.

\subsubsection{Effect of the Earth's rotation}

To discuss the diurnal frequency modulation it is convenient 
to rewrite the first term on the RHS of Eq. (\ref{Df}) as
\begin{equation}
{1 \over 2 c^2}(\vec{v}_{\oplus i} - \vec{v}_{\rm rot})^2
= {1 \over 2 c^2}(v_{\oplus i}^2 + v_{\rm rot}^2 
- 2 \vec{v}_{\oplus i} \cdot \vec{v}_{\rm rot})
\label{ko}
\end{equation}
where
\begin{equation}
\vec{v}_{\oplus i} = \vec{v}_{\odot i} - \vec{v}_{\rm orb}
\label{vo}
\end{equation} 
is the flow velocity in a reference frame which is co-moving 
but not co-rotating with the Earth, and 
\begin{equation}
\vec{v}_{\rm rot} = 0.465 {{\rm km} \over {\rm s}}
(- \sin \phi_s~\hat{x}_\oplus + \cos \phi_s~\hat{y}_\oplus) \cos l
\label{vs}
\end{equation}
is the velocity of the laboratory in that reference frame.  We 
use Earth based coordinates $(x_\oplus, y_\oplus, z_\oplus)$ 
such that $\hat {z}_\oplus$ is the Earth's rotation axis pointing 
North, $\hat {x}_\oplus$ points towards the Vernal Equinox (the 
direction of the Sun at the time of the vernal equinox) 
and $\hat {y}_\oplus = \hat{z}_\oplus \times \hat{x}_\oplus$.
$l$ is the latitude of the laboratory, $\phi_s \equiv 2 \pi {t \over t_s}$ 
where $t$ is sidereal time - i.e. the time elapsed since the Vernal
Equinox last crossed the local meridian - and $t_s$ is the sidereal day.  
Eq. (\ref{vs}) neglects the asphericity of the Earth.  The second term 
in Eq. (\ref{ko}) increases all peak frequencies by a tiny time-independent 
amount
\begin{equation}
{\Delta f_i \over \nu_a}\Bigg|_{\oplus~{\rm rot}} 
= {v_{\rm rot}^2 \over 2 c^2} =
1.2~10^{-12}~\cos^2 l~~~~\ .
\label{ss}
\end{equation}
The last term in Eq. (\ref{ko}) can be spelled out:
\begin{equation}
- {1 \over c^2} \vec{v}_{\oplus i} \cdot \vec{v}_{\rm rot} = 
0.465~{{\rm km} \over {\rm s}}~\cos l~(\sin \phi_s~v_{\oplus i x}
- \cos \phi_s~v_{\oplus i y}){1 \over c^2}~~~~\ .
\label{sp}
\end{equation}
Since $v_{\oplus i}$ is of order 300 km/s, the daily frequency 
modulation amplitude is of order $10^{-9}$.  Eq. (\ref{sp}) shows 
that the measurement of the amplitude and the phase of the daily 
frequency modulation determines the values of $v_{\oplus i x}$ and 
$v_{\oplus i y}$ on any day of the year.  Moreover, the magnitude of 
$v_{\oplus i z}$ is determined by measuring the day-averaged frequency 
shift away from $\nu_a$:
\begin{equation}
{\Delta f_i \over \nu_a}\Bigg|_{{\rm day} \atop {\rm averaged}} = 
{1 \over 2 c^2}(v_{\oplus i x}^2 + v_{\oplus i y}^2 + 
v_{\oplus i z}^2) + {G M_\oplus \over c^2 R_\oplus}~~~\ .
\label{da}
\end{equation}
The ambiguity in the sign of $v_{\oplus i z}$ is removed by 
observing the annual modulation; see below.  Thus, on any given 
day, all three components of $\vec{v}_{\oplus i}$ can be measured.

\subsubsection{Effect of the Sun's gravity and of the Earth's 
orbital motion}

Next we consider the annual frequency modulation, i.e. the time 
dependence of 
\begin{equation}
{\Delta f_i \over \nu_a}\Bigg|_{{\rm day} \atop {\rm averaged}}
- {G M_\oplus \over c^2 R_\oplus} =
{1 \over 2 c^2} (\vec{v}_{\odot i}- \vec{v}_{\rm orb})^2~~~\ .
\label{an}
\end{equation}
Using energy conservation for the dark matter particles in the flow
we have
\begin{equation}
{\Delta f_i \over \nu_a}\Bigg|_{{\rm day} \atop {\rm averaged}} 
- {G M_\oplus \over c^2 R_\oplus} =
{v_0^2 \over 2 c^2} + {G M_\odot \over c^2 r(t)} 
+ {1 \over 2 c^2}v_{\rm orb}^2(t) - 
{1 \over c^2} \vec{v}_{\odot i}\cdot \vec{v}_{\rm orb}~~~\ ,
\label{ane}
\end{equation}
where $t$ now stands for time of year.  Furthermore, we may use 
energy conservation in the orbital motion of the Earth
\begin{equation}
{1 \over 2}v_{\rm orb}^2(t) - {G M_\odot \over r(t)} 
= - {G M_\odot \over 2 a_\oplus}
\label{eo}
\end{equation}
to rewrite Eq. (\ref{ane}) as
\begin{equation}
{\Delta f_i \over \nu_a}\Bigg|_{{\rm day} \atop {\rm averaged}} 
- {G M_\oplus \over c^2 R_\oplus} =
{v_0^2 \over 2 c^2} + {G M_\odot \over c^2}\left({2 \over r(t)} - 
{1 \over 2 a_\oplus}\right) - {1 \over c^2}\vec{v}_{\odot i}
\cdot \vec{v}_{\rm orb}~~~\ . 
\label{an2}
\end{equation}
Here, $a_\oplus = 1 {\rm A.U.} = 1.4960~10^{13}$ cm is the semi-major
axis of the Earth's orbit.

Let $(r, \Theta, \Phi)$ be spherical coordinates centered on the Sun.
The polar angle $\Theta$ is measured relative to $\hat{Z}$, the 
direction perpendicular to the plane of the Earth's orbit, and $\Phi = 0$
in the direction of the Vernal Equinox, i.e. the direction of the Sun 
as viewed from Earth at the time of the vernal equinox. Thus the Earth is
at $\Phi = 0$ at the time of the autumnal equinox (near Sept. 23). The 
angular position $\Phi(t)$ of the Earth and its distance $r(t)$ to the 
Sun are given by the parametric equations: 
\begin{eqnarray} \tan
({\Phi - \Phi_p \over 2}) &=& \sqrt{1 + e_\oplus \over 1 - e_\oplus} 
\tan{\psi \over 2} \nonumber\\ 
r &=& a_\oplus (1 - e_\oplus \cos\psi)\nonumber\\ 
t - t_p &=& \sqrt{a_\oplus^3 \over G M_\odot} 
(\psi - e_\oplus \sin \psi) 
\label{KN} 
\end{eqnarray} 
where $e_\oplus = 0.0167$ is the eccentricity of the Earth's orbit and
$\psi$ is the 'eccentric anomaly' parameter.  In Eqs.(\ref{KN}) the
subscript $p$ indicates that the quantity is evaluated at perihelion,
which occurs near Jan. 4.  Using Eqs. (\ref{KN}), the second term on 
the RHS of Eq. (\ref{an2}) may be rewritten as 
\begin{equation} 
{G M_\odot \over c^2 a_\oplus} \left[{3\over 2} 
+ 2 e_\oplus \cos \left({2 \pi (t - t_p) \over {\rm year}}\right) +
0(e_\oplus^2)\right] 
\label{or} 
\end{equation} with
\begin{equation} 
{G M_\odot \over c^2 a_\oplus} = 0.987~10^{-8}~~~~\ .
\label{su} 
\end{equation}
Eq. (\ref{or}) describes an increase of all frequencies by the same
amount, the time-dependent part of which is of order $10^{-10}$.

Finally we analyze the last term in Eq. (\ref{an2}).  Although we 
kept it for last, it is by far the largest contribution to the 
frequency modulation.  For the sake of clarity, we first neglect 
the eccentricity $e_\oplus$ of the Earth's orbit.  

\subsubsection{Neglecting the eccentricity of the Earth's orbit}

In the $e_\oplus = 0$ limit, $\vec{v}_{\rm orb} = 
\sqrt{G M_\odot \over a_\oplus} \hat{\Phi}$.  Combining this with 
Eq. (\ref{vi}) we have for all flows $(i =1,2,3)$
\begin{equation}
- {1 \over c^2} \vec{v}_{\odot i} \cdot \vec{v}_{\rm orb} =
+ \hat{\theta} \cdot \hat{\Phi} {v_0 \over c^2} 
\sqrt{G M_\odot \over a_\oplus^3} b_i~~~~,
\label{a0}
\end{equation}
since $\hat{\Phi} \cdot \hat{r} = 0$. Let $(\Theta_0, \Phi_0)$ be 
the polar coordinates of the flow velocity $\vec{v}_0 = v_0 \hat{z}$ 
in the Sun centered coordinate frame defined above.  We have then
(for $\rho > 0$)
\begin{eqnarray}
\hat{\theta} \cdot \hat{\Phi} &=&
{\hat{r} \cos \theta - \hat{z} \over \sin \theta} \cdot \hat{\Phi}
= - {\hat{z} \cdot \hat{\Phi} \over \sin \theta} \nonumber\\
&=& + {\sin \Theta_0 \sin(\Phi - \Phi_0) \over 
\sqrt{1 - \sin^2 \Theta_0 \cos^2 (\Phi - \Phi_0)}}~~~\ .
\label{a02}
\end{eqnarray}
In terms of $a = {G M_\odot \over v_0^2}$, we have therefore
\begin{equation}
- {1 \over c^2} \vec{v}_{\odot i} \cdot \vec{v}_{\rm orb} =
+ {v_0^2 \over c^2} \sqrt{a \over a_\oplus} {b_i \over a_\oplus}
{\sin \Theta_0 \sin(\Phi - \Phi_0) \over  
\sqrt{1 - \sin^2 \Theta_0 \cos^2 (\Phi - \Phi_0)}}~~~~\ ,
\label{a03}
\end{equation}
where $\Phi = 2 \pi {t \over {\rm year}}$, and $t$ is time since the 
last autumnal equinox.

Let us estimate the frequency modulation amplitudes predicted by 
Eq. (\ref{a03}).  For flow~1, $|b_1| \sim a_\oplus$ generically, 
i.e. away from the spike caustic.  Since
\begin{equation}
a \equiv {G M_\odot \over v_0^2} = 1.4746~10^{11}~{\rm cm}~
\left({300 {\rm km/s} \over v_0}\right)^2~~~~ ,
\label{a}
\end{equation}
the frequency modulation amplitude is typically of order 
$10^{-7}$ for flow 1.  For flow 2 [see Eq. (\ref{b12})],
$|b_2| \sim a \sim 10^{-2} a_\oplus$ generically, i.e. 
away from the spike.  Hence the frequency modulation 
amplitude is typically of order $10^{-9}$ for flow 2.
Close to the spike, flows 1 and 2 are alike: 
$b_{1 \atop 2} \simeq \pm b_{\rm spike}$ with 
\begin{equation}
b_{\rm spike} \equiv \sqrt{2 a a_\oplus} = 
2.1005~10^{12}~{\rm cm}~\left({300 {\rm km/s} \over v_0}\right)~~~\ .
\label{bs}
\end{equation}
Therefore, near the spike, the frequency shifts for flows 1 and 2 
are both of order $10^{-8}$.  For flow 3, $b_3 \sim R_\odot$ 
everywhere, and its frequency modulation amplitude is 
therefore of order $10^{-9}$.

Let us consider more closely the frequency shifts of flows 
1 and 2 when the Earth comes near the spike caustic, i.e.
when $Y = {2a \over r \sin^2 ({\theta \over 2})} >> 1$.  For 
the Earth to come near the spike, the unperturbed flow direction
$\hat{z}$ must lie close to the ecliptic plane.  So let 
$\Theta_0 = {\pi \over 2} - \delta$, and $\Phi - \Phi_0 = \epsilon$,
with $\delta, \epsilon << 1$.  Then
\begin{equation}
- {1 \over c^2} \vec{v}_{\odot {1 \atop 2}} \cdot \vec{v}_{\rm orb}
= \pm \sqrt{2} {G M_\odot \over c^2 a_\oplus}
{\epsilon \over \sqrt{\delta^2 + \epsilon^2}}~~~ ,
\label{as}
\end{equation}
where we used Eqs. (\ref{a03}) and (\ref{bs}).

\subsubsection{Including the eccentricity of the Earth's orbit}

Finally, let us include the corrections due to the finite eccentricity
of the Earth's orbit.  Using Eqs. (\ref{KN}), we have:
\begin{equation}
\vec{v}_{\rm orb} = \sqrt{G M_\odot \over a_\oplus}
{1 \over 1 - e_\oplus \cos \psi} \left[e_\oplus \sin \psi~\hat{r}
+ \sqrt{1 - e_\oplus^2}~\hat{\Phi}\right]~~~~\ .
\label{vor}
\end{equation}
Combining this with Eq. (\ref{vi}), we find ($i = 1,2,3$)
\begin{eqnarray}
- {1 \over c^2} \vec{v}_{\odot i} \cdot \vec{v}_{\rm orb}
= - {v_0^2 \over c^2} {\sqrt{a a_\oplus} \over r}
[&\pm& e_\oplus \sin \psi \sqrt{1 + {2a \over r} - ({b_i \over r})^2}
\nonumber\\
&-& \sqrt{1 - e_\oplus^2} {b_i \over r} {\sin \Theta_0 \sin (\Phi - \Phi_0)
\over \sqrt{1 - \sin^2 \Theta_0 \cos^2 (\Phi - \Phi_0)}}]~~\ .
\label{mf}
\end{eqnarray}
In Eq. (\ref{mf}) the $\pm$ sign corresponds to flows with 
$\dot{r}~{> \atop <}~0$.  Flows 2 and 3 always have $\dot{r} > 0$.
$\Phi,~\psi$ and $r$ are given as functions of time $t$ by 
Eqs. (\ref{KN}).  The impact parameters $b_i$ are given in
Eqs. (\ref{b12}) and (\ref{b23}), with $z = r \cos \theta$, 
$\rho = r \sin \theta$, and 
\begin{equation}
\cos \theta = \sin \Theta_0 \cos (\Phi - \Phi_0)~~~\ .
\label{th}
\end{equation}

\subsubsection{Frequency shifts on the measurement time scale}

Finally, we estimate the frequency shift of a peak in the axion signal 
in a 100 sec time interval.  This is of order the measurement 
integration time of the cavity detector in search mode.  The 
shift due to the Earth's orbital motion is at most of order 
\begin{equation}
\Delta f|_{orb \atop 100{\rm sec}} 
\lsim 10^{-7} f {2 \pi \over {\rm year}} (100~{\rm sec}) \simeq
2 \cdot 10^{-12} f
\label{os}
\end{equation}
for flow 1, less for the others.  The shift due to the Earth's rotation
\begin{equation}
\Delta f|_{rot \atop 100{\rm sec}} \lsim 
1.5 \cdot 10^{-9} {2 \pi \over {\rm day}}(100~{\rm sec}) \simeq
10^{-11}f ~~~\ .
\label{rs}
\end{equation}
We see that the effect of Earth's rotation is larger than the effect
of its orbital motion on this short time scale.  In the ADMX experiment, 
$f \sim 1$ GHz and $t \sim 100$ sec \cite{ADMX}, so that 
$\Delta f \lsim 10$ mHz, and hence the frequency resolution 
$\delta f = {1 \over t} \sim 10$ mHz is as high as is profitable 
in search mode.

\subsection{Density modulation}

The size of a peak in the axion signal is proportional to the 
density of the corresponding flow at the detector location.  
A peak's size is much less accurately measured than its 
frequency.  So we neglect in this subsection the subleading
effects associated with the Earth's gravity and rotation, 
and the eccentricity of the Earth's orbit.

The density of the flows that did not go through the Sun
is given by Eqs. (\ref{d12}) and (\ref{Y}), with $r = a_\oplus$, 
$\Phi = 2 \pi {t \over {\rm year}}$, and $\theta$ given by 
Eq. (\ref{th}).  Away from the spike, i.e. for $Y << 1$, 
\begin{eqnarray}
d_1 - d_0 = d_2 = d_0~({1 \over 16} Y^2 + 0(Y^3)) 
\simeq {d_0 \over 4} \left({a \over a_\oplus}\right)^2 
{1 \over \sin^4 ({\theta \over 2})} \nonumber\\
\simeq d_0~2.43~10^{-5}~\left({300 {\rm km/s} \over v_0}\right)^2~
{1 \over \sin^4 ({\theta \over 2})}~~~\ .
\label{dm12}
\end{eqnarray}
The above approximation breaks down when $Y \gsim 1$ or
\begin{equation}
\theta \lsim 16^\circ \left({300~{\rm km/s} \over v_0}\right)~~~~\ .
\label{dmb}
\end{equation}
Close to the spike ($Y >> 1$), the densities of flows 1 and 2 
are large and nearly equal:
\begin{equation}
d_1 \simeq d_2 \simeq {d_0 \over 4} \sqrt{Y} \simeq 
d_0 \sqrt{a a_\oplus \over 2} {1 \over \rho} \simeq
d_0~\sqrt{a \over a_\oplus} {1 \over \sqrt{2(\delta^2 + \epsilon^2)}}
\label{ns}
\end{equation}
with $\delta = {\pi \over 2} - \Theta_0$ and $\epsilon = \Phi - \Phi_0$.
Note that we are assuming here that the spike is truly a {\it line}
caustic.  The spike is a line caustic if the incoming flow is uniform and
the Sun is spherically symmetric.  However, a line caustic is a degenerate
case.  Generically, dark matter caustics are surfaces.  When observed 
with sufficient resolution, the central line of the spike will be seen 
to be a caustic surface whose properties depend on the asphericity 
of the Sun and the lack of uniformity of the incoming flow.

The density of the flows that did go through the Sun are given 
by Eq. (\ref{d23}).  Let us first consider the density of flow 3 
downstream of the Sun.  Near $\theta = 0$,  $b_3(\theta) = - B \theta$
where $B$ depends on the mass distribution inside the Sun; see Fig. 2.  
Hence
\begin{equation}
d_3(\theta) \simeq d_0 \left({B \over a_\oplus}\right)^2
\label{d30}
\end{equation}
for $\theta << 1$.  For the model considered in ref. \cite{SW}, 
$B \sim 0.7 R_\odot$ for $v_0 = 300$ km/s, and hence 
$d_3 \sim 10^{-5} d_0$.   Generically, i.e. away from the spike, 
$d_2$ and $d_3$ have the same order of magnitude.

Finally we consider the density of flows 2 and 3 near the skirt 
caustic.  Let $b_{\rm c}$ be the value of the impact parameter 
$b_{\rm s}$ for which the scattering angle $\Theta$ is maximum.  
For $\Theta$ near $\Theta_{\rm max}$
\begin{equation}
\Theta(b_{\rm s}) = \Theta_{\rm max} - {1 \over 2} H 
(b_{\rm s} - b_{\rm c})^2
\label{Thb}
\end{equation}
with 
\begin{equation}
H = - {d^2 \Theta \over d b_{\rm s}^2} \bigg|_{b_{\rm s} = b_{\rm c}} 
> 0~~~\ .
\label{H}
\end{equation}
Inserting this into Eq. (\ref{d23}) we have 
\begin{equation}
d_2 \simeq d_3 \simeq {d_0 b_{\rm c} \over
r^2 \sin \Theta_{\rm max} \sqrt{2 H (\Theta_{\rm max} - \theta)}}
\label{skc}
\end{equation}
for $\theta \leq \Theta_{\rm max}$.  For $\theta > \Theta_{\rm max}$,
$d_2$ and $d_3$ vanish.  The densities of flows 2 and 3 diverge 
for $\theta \rightarrow \Theta_{{\rm max}, -}$ as the inverse square 
root of distance to the caustic surface.  This behavior is characteristic 
of a simple fold catastrophe.  However, the coefficient of the 
singularity is inversely proportional to $r^2$.  At the Earth's 
distance to the Sun, the skirt caustic is quite feeble, as we will see.
   
The values of $b_{\rm c}$, $\Theta_{\rm max}$ and $H$ depend on the 
solar mass distribution and on $v_0$.  For the solar mass 
distribution considered in ref. \cite{SW} and $v_0 = 300$ km/s, 
one has $b_{\rm c} \simeq 1.35 R_\odot$, $\Theta_{\rm max} \simeq
107^\circ$ and $H \simeq {200 \over R_\odot^2}$, in which case
\begin{equation}
d_2 \simeq d_3 \simeq d_0 {1.53 ~10^{-6} \over 
\sqrt{\Theta_{\rm max} - \theta}}
\label{fe}
\end{equation}
for $r = a_\oplus$.  In view of Eq. (\ref{th}), the Earth 
crosses a skirt caustic twice per year if
\begin{equation}
|\cos \Theta_{\rm max}| \leq \sin \Theta_0~~~~\ .
\label{cr}
\end{equation}
The skirt crossing times are $t_{\rm sk} = \Phi_{\rm sk} 
{{\rm year} \over 2 \pi}$ where $\Phi_{\rm sk}$ are the
solutions of
\begin{equation}
\cos \Theta_{\rm max} = \sin \Theta_0 \cos(\Phi_{\rm sk} - \Phi_0)~~~\ .
\label{Tsk}
\end{equation}
If $\sin \Theta_0 < |\cos \Theta_{\rm max}|$, the Earth never crosses 
the skirt caustic.  In the latter case, flows 2 and 3 are always 
absent (present) on Earth if $\cos \Theta_{\rm max} >~(<)~0$.  As 
a function of time
\begin{equation}
d_{2,3}(t) \simeq {d_0 b_{\rm c} \over \sqrt{H} a_\oplus^2}
{1.58~10^3 \over \sqrt{\sin \Theta_0~~\sin \Theta_{\rm max}~~
|\sin (\Phi_{\rm sk} - \Phi_0)|}} 
\sqrt{{\rm sec} \over |t - t_{\rm sk}|}~~~~,
\label{skt}
\end{equation}
near the time of caustic crossing, on the side with two extra flows. 
For some flows the orientation of the Earth's orbit may be such
that $|\Theta_{0} \pm \Theta_{\rm{max}}| \simeq \pi/2$ in which
case the Earth's trajectory grazes the skirt caustic, and the
density enhancement lasts a relatively long time.

\section{WIMP Signal Modulation}

In this section, we review briefly the equations describing WIMP
detection in the laboratory, derive the signal modulation for a 
single flow with zero velocity dispersion, and discuss the 
implications for the caustic ring halo model.

\subsection{General formalism}

The experimental approach to WIMP detection in the laboratory 
is to measure the recoil energy $E_r$ of nuclei off of which 
WIMPs have scattered elastically \cite{Wdet}.  See ref. 
\cite{Wrev} for reviews.  Although it is highly desirable 
to measure the recoil direction as well, this capability has 
not yet been achieved in present detectors.

The differential event rate per unit detector mass is
\beq
dR=\frac{d}{m_\chi m_N} \; v f(v) \; 
\frac{d \sigma}{d \abs{\vec{q}}^2} \;
d \abs{\vec{q}}^2 \; dv
\eeq
where $m_\chi$ is the WIMP mass, $d$ the local WIMP mass density, 
$m_N$ the target nucleus mass, $f(v)$ the normalized CDM velocity 
distribution in the laboratory reference frame, $\vec{q}$ the 
momentum transfer, and $d \sigma/d \abs{\vec{q}}^2$ the differential
cross-section for elastic WIMP-nucleus scattering.  The recoil 
energy deposited in the detector is
\beq
E_r=\frac{\abs{\vec{q}}^2}{2 m_N} \qquad \ .
\eeq
For given initial WIMP velocity $v$, the maximum momentum transfer
(and hence maximum recoil energy) occurs when the scattering is 
back-to-back in the center of mass frame, in which case 
$\abs{\vec{q}} = \abs{\vec{q}}_{\rm max} = 2 m_r v$, where
$m_r = {m_N m_\chi \over m_N + m_\chi}$ is the reduced mass.
The differential cross-section is usually expressed in terms
of a ``standard" cross-section at zero-momentum transfer, called 
$\sigma_0$, and a form factor $F(\abs{\vec{q}})$, normalized 
so that $F(0) = 1$:
\beq
\frac{d \sigma}{d \abs{\vec{q}}^2}=
\frac{\sigma _0}{4 m_r^2 v^2} \; F(\abs{\vec{q}})^2~~~\ .
\eeq
The model-dependent factors describing the interactions of a 
specific WIMP enter $\sigma _0$, while the nuclear effects 
are encoded in $F(\abs{\vec{q}})$.

Integrating over the WIMP velocity distribution, we obtain 
the event rate per unit recoil energy:
\begin{eqnarray}
\frac{dR}{dE_r} &=& \frac{\sigma _0 d}{2 m_\chi m_r^2} \;
F(\abs{\vec{q}})^2 \; \int _{v_{\rm min}}^\infty 
\frac{f(v)}{v} \; dv \qquad \nonumber\\
&=& {\sigma_0 d \over \sqrt{\pi} m_\chi m_r^2}
F(\abs{\vec{q}})^2 {T(E_r) \over 220~{\rm km/s}}~~~ ,
\end{eqnarray}
where $\abs{\vec{q}} = \sqrt{2 m_N E_r}$, and 
$v_{\rm min} = {\abs{\vec{q}} \over 2 m_r}$ is the minimum
velocity needed to transfer a given energy $E_r$.  $T(E_r)$
is a dimensionless function containing all relevant information 
about the velocity distribution.  It is conventionally defined as
\beq
T(E_r) = {\sqrt{\pi} \over 2} (220~{\rm km/s})
\int _{v_{\rm min}}^\infty \frac{f(v)}{v} \; dv ~~~\ .
\label{TE}
\eeq
Finally, the total event rate within an observed range of recoil 
energies is
\beq
R=\int_{E_1}^{E_2} \frac{dR}{dE_r} \; dE_r \qquad 
\eeq
where $E_1$ and $E_2$ are respectively the minimum and 
maximum values of that range.

\subsection{The single flow plateau} 

Consider a single cold (i.e. zero velocity dispersion) flow, 
of density $d_F$ and speed $v_F$ relative to the laboratory.
In that case
\beq
f(v)=\delta(v-v_F)
\eeq
and
\beq
\frac{dR}{dE_r}=\frac{\sigma _0 \, d_F}{2 m_\chi m_r^2 \, v_F} \;
F^2(\abs{\vec{q}}) \; H(v_F-v_{min})
\label{diffrate}
\eeq
where $H(x)$ is the Heaviside step function.  In the limit
where the form factor is slowly varying over the relevant 
momentum transfer range, the contribution of a single cold 
flow to the recoil energy spectrum is a flat plateau \cite
{ver,gre,gel,stream}:
\begin{eqnarray}
{dR \over dE_r}~&=&~{\sigma_0 d_F \over 2 m_\chi m_r^2} {F^2 \over v_F}
~~~~~~~~~~~~~~~~~~{\rm for}~~~E_r < E_{\rm max}\nonumber\\
&=&~~~0~~~~~~~~~~~~~~~~~~~~~~~~~~~~~{\rm for}~~~E_r > E_{\rm max}
\label{diff2}
\end{eqnarray}
where
\beq
E_{\rm max} = {2 m_r^2 v_F^2 \over m_N}
\label{Emax}
\eeq
is the maximum recoil energy for the flow.  The total rate within 
an observed range of recoil energies, $E_1 < E_r < E_2$, is 
\beq
R=\frac{\sigma _0 \, d_F}{2 m_\chi m_r^2 \, v_F} \;
\left[ {\cal F} \left( \min \{ E_2, E_{\rm max} \} \right) -
{\cal F} \left( \min \{ E_1, E_{\rm max} \} \right) \right]
\eeq
where
\beq
{\cal F}(E)=\int _0^E F^2(\abs{\vec{q}}) \; dE_r \qquad .
\eeq
For $E_1 = 0$ and $E_2 = \infty$, we recover the usual expression 
for the total rate (per unit detector mass)
\beq
R = \sigma_{\rm tot} v_F {d_F \over m_\chi m_N}
\eeq
where
\beq
\sigma_{\rm tot} = \sigma_0 {1 \over E_{\rm max}}
\int_0^{E_{\rm max}} F(\abs{\vec{q}})^2~dE_r
\eeq
is the total cross-section.  

The differential rate ${dR \over dE_r}$ as a function of 
$E_r$ is a plateau whose height is inversely proportional 
to the flow velocity $v_F$, and whose edge is proportional 
to $v_F^2$.  So the total rate increases with increasing 
flow velocity proportionately to $v_F$, as one would expect, 
but the rate within an observed recoil energy range 
($E_1 < E_r < E_2$) decreases with increasing flow 
velocity if $E_2 < E_{\rm max}$.

\subsection{Signal modulation for a single flow}

As discussed in Section II, a single cold homogeneous flow of density
$d_0$ and velocity $\vec{v}_0$ far upstream of the Sun, produces in the
vicinity of the Sun one or three flows, depending on location.  Formulas
for the densities $d_i(\vec{r})$ and velocities $\vec{v}_{\odot i}(\vec{r})$ 
of the flows ($i=1$ or $i=1,2,3$) were given in Section II for all
$\vec{r}$.  The flow densities $d_i(t)$ and speeds $v_i(t)$ on Earth 
as functions of time of day and of year were obtained in Section
III.  (The latter section is motivated by the axion dark matter
experiment.  However the formulas can be applied to WIMPs as well.)  
To obtain the time-dependent signal ${dR \over dE_r}(t)$ caused
by a single mother flow, one must therefore replace $d_F$ with 
$d_i(t)$ and $v_F$ with $v_i(t)$ in Eq (\ref{diffrate}), and sum 
over $i$.

The expressions for $d_i(t)$ and $v_i(t)$ in Section III, however, include
many corrections to which present WIMP detectors are not sensitive. In
particular, flow 3 is unlikely to be observed in present WIMP detectors
because its density $d_3$ is of order $10^{-5} d_0$. Likewise, flow 2 is
unlikely to be observed except when the Earth approaches the spike
caustic.  The most important effect for WIMP detectors is the annual
modulation of the velocity $v_1$ of flow 1 due to the Earth's orbital
motion.  It is of order 10\%.  A short list of the lesser effects is as
follows.  The Earth's gravity increases all the $v_i$ by a constant amount
of order 0.1\%. The Earth's rotation produces a diurnal modulation of the
$v_i$ of order 0.3\%.  The effect of the Sun's gravity on the $v_i$ is of
order 1\%.  The effect of the Sun's gravity on the $d_i$ is of order
$10^{-4}$ except near the spike caustic where $d_1$ and $d_2$ diverge.  
The corrections due to the eccentricity of the Earth's orbit are of order
0.1\%.

Here, we recapitulate our results for $d_i(t)$ and $v_i(t)$ including 
all effects down to the 0.3\% level of the diurnal modulation.  So we
neglect the eccentricity $e_\oplus$ of the Earth's orbit. We also neglect 
the existence of the third flow and all other effects, such as the 
skirt caustic, associated with the finite size of the Sun.  With these 
approximations, the densities are given by 
Eq. (\ref{d12}) with 
\beq
Y = 4\left({v_{\rm orb} \over v_0}\right)^2~{1 \over
1~-~\sin\Theta_0~\cos(\Phi - \Phi_0)}~~~\ .
\label{YY}
\eeq
We used ${a \over r} = {a \over a_\oplus} = ({v_{\rm orb} \over v_0})^2$ 
since the eccentricity of the Earth's orbit is neglected.  As before 
$(\Theta_0, \Phi_0)$ are the solar polar coordinates of the initial flow 
direction $\hat{v}_0$, and $\Phi = 2 \pi {t \over {\rm year}}$ where $t$ is 
time since the autumnal equinox (near Sept. 21).  The 
flow speeds on Earth are
\beq
v_i = \sqrt{(\vec{v}_{\odot i} - \vec{v}_{\rm orb}
- \vec{v}_{\rm rot})^2}
\eeq
where $\vec{v}_{\odot i} = \vec{v}_0 + \Delta \vec{v}_{\odot i}$
is the flow velocity corrected for the effect of solar gravity, 
and $\vec{v}_{\rm orb}$ and $\vec{v}_{\rm rot}$ are the 
velocity components the laboratory has in the rest frame of 
the Sun due to the Earth's orbital and rotational motions.
We have in solar coordinates
\begin{eqnarray}
\vec{v}_{\rm orb} &=& v_{\rm orb}~ 
(- \sin \Phi ~\hat{X} + \cos \Phi~\hat{Y})\nonumber\\
\vec{v}_{\rm rot} &=& v_{\rm rot} \cos l~[- \sin \phi_s~\hat{X}
+ \cos \phi_s (\cos \gamma~\hat {Y}~-~\sin \gamma~\hat{Z})] \nonumber\\
\vec{v}_{\odot {1 \atop 2}} &=& {1 \over 2} v_0 (1 \pm \sqrt{1+Y})~
[\sin \Theta_0(\cos \Phi_0~\hat{X}~+~\sin\Phi_0~\hat{Y})
+ \cos \Theta_0~\hat{Z}]\nonumber\\
&+& {1 \over 2} v_0 (1 \mp \sqrt{1+Y})~
(\cos \Phi~\hat{X}~+~\sin \Phi~\hat{Y})~~~\ ,
\end{eqnarray}
where $v_{\rm orb} = 29.8$ km/s and $v_{\rm rot} = 0.46$ km/s.
$l$ and $\phi_s$ were defined in Section III, after Eq. (3.10).
$\gamma = 23.44^\circ$ is the angle between the Earth's equatorial 
plane and its orbital plane.  The latter is usually referred to as 
'the ecliptic'.

The most important effect is caused by the Earth's orbital motion.
If we neglect the lesser effects of solar gravity and terrestrial 
rotation, we have $d_1 = d_0$, $d_2 = 0$, and 
\begin{eqnarray}
v_1 &=& \sqrt{(\vec{v}_0~-~\vec{v}_{\rm orb})^2} \simeq 
v_0~-~\hat{z}\cdot\vec{v}_{\rm orb}\nonumber\\
&\simeq& v_0~+~v_{\rm orb}~\sin \Theta_0~\sin (\Phi - \Phi_0)~~~\ .
\end{eqnarray}
At this order, the speed modulation is sinusoidal with 
amplitude ${v_{\rm orb} \over v_0} \sin \Theta_0 
= 10\%~({300~{\rm km/s} \over v_0}) \sin \Theta_0$.  The 
modulation amplitude is largest when the initial flow 
direction is near the ecliptic plane.  In that case, however, 
the spike caustic is near the ecliptic plane as well and the 
effect of solar gravity is important when the Earth approaches 
the spike caustic.

If we include the effect of solar gravity, but still neglect 
that of Earth's rotation, we find
\beq
(v_{1 \atop 2})^2 = v_0^2~+~3 v_{\rm orb}^2~-~
\vec{v}_0 \cdot \vec{v}_{\rm orb}~(1~\pm~\sqrt{1+Y})~~~ ,
\eeq
where we used $v_{\rm orb}^2 = {G M_\odot \over a_\oplus}$.
Solar gravity increases all flow speeds by the relative amount 
$({v_{\rm orb} \over v_0})^2 = 1\%~({300~{\rm km/s} \over v_0})^2$
and modifies the $\hat{z} \cdot \vec{v}_{\rm orb}$ term in a 
position-dependent way. When approaching the spike, $Y$ diverges. 
However, $v_i$ remains finite because $\vec{v}_0 \cdot
\vec{v}_{\rm orb} \rightarrow 0$ there.  On the other hand, as
Eq. (\ref{d12}) shows, both  $d_1$ and $d_2$ diverge at the spike.  
Setting $\Theta_0 = {\pi \over 2} - \delta$ and 
$\Phi = \Phi_0 + \epsilon$, we have
\begin{eqnarray}
Y &\simeq& 8 \left({v_{\rm orb} \over v_0}\right)^2 
{1 \over \delta^2 + \epsilon^2}\nonumber\\
(v_{1 \atop 2})^2 &\simeq& v_0^2~+~v_{\rm orb}^2~
(3~\pm~{2\sqrt{2}\epsilon \over \sqrt{\epsilon^2 + \delta^2}})\nonumber\\
d_1 \simeq d_2 &\simeq& d_0 \left({v_{\rm orb} \over v_0}\right)
{1 \over \sqrt{2(\epsilon^2 + \delta^2)}}
\end{eqnarray}
for $\epsilon, \delta << 0$.  The maximum value of the density 
enhancement $F \equiv {1 \over d_0}(d_1 + d_2)$ is
\beq
F_{\rm max} \simeq \sqrt{2}~{v_{\rm orb} \over v_0}~ 
{1 \over \delta}~~~\ .
\eeq
The duration of a peak with enhancement factor larger than $F$ is
\beq
\Delta t = {{\rm year} \over \pi}~\delta~ 
\sqrt{({F_{\rm max} \over F})^2 -1}
\eeq
for $1 << F \leq F_{\rm max}$.  

Finally, including the effect of Earth's rotation, we have
\begin{eqnarray}
v_i^2 &=& (\vec{v}_{\odot i} - \vec{v}_{\rm orb})^2~+~
{2 G M_\odot \over a_\oplus} - 
2~(\vec{v}_{\odot i} - \vec{v}_{\rm orb})\cdot\vec{v}_{\rm rot}\nonumber\\
&=& v_0^2~+~3~v_{\rm orb}^2~-~
\vec{v}_0 \cdot \vec{v}_{\rm orb} (1~\pm~\sqrt{1+Y})
~-~ 2~\vec{v}_{\odot i} \cdot \vec{v}_{\rm rot} 
~+~ 2~\vec{v}_{\rm orb} \cdot \vec{v}_{\rm rot}~~~\ .
\label{vi2}
\end{eqnarray}
The last term 
\beq
2~\vec{v}_{\rm orb} \cdot \vec{v}_{\rm rot} =
2~v_{\rm orb} v_{\rm rot} \cos l~[\cos \gamma~\cos \Phi~\cos \phi_s
~+~\sin \Phi~\sin \phi_s]
\eeq
is a correction of order $10^{-3}$. The penultimate term in
Eq. (\ref{vi2}) is 
\begin{eqnarray}
- 2 \vec{v}_{\odot i} \cdot \vec{v}_{\rm rot} = 
&-& v_0 v_{\rm rot} \cos l~[(1~\pm~\sqrt{1+Y}) 
(- \sin \phi_s~\sin \Theta_0~\cos \Phi_0~\nonumber\\
&+&~\cos\gamma~\cos\phi_s~\sin\Theta_0~\sin\Phi_0~-~
\sin\gamma~\cos\phi_s~\cos\Theta_0)\nonumber\\
~&+&~(1~\mp~\sqrt{1+Y})(- \sin\phi_s~\cos\Phi~+~
\cos\gamma~\cos\phi_s~\sin\Phi)]
\end{eqnarray}
with $Y$ given by Eq. (\ref{YY}).  

\subsection{WIMP signal modulation in the Caustic Ring Halo Model}

The caustic ring halo model predicts a set of discrete flows in the solar
neighborhood.  The model is briefly described in Appendix B. The list of
flows is given in Table I in galactic coordinates, and in Table II in
solar coordinates.  The local dark matter distribution is dominated by one
pair of flows.  They are associated with the caustic ring of dark matter
nearest to us, and are given by the $n=5$ entries of the Tables.  One of
the $n=5$ flows, dubbed the 'big flow', has density of order three times
the sum of all the other flows combined.  This feature simplifies the
analysis of the model's predictions for the WIMP signal modulation.  One
complicating factor is our present ignorance whether the velocity of the
big flow is $\vec{v}_5^-$, in which case the velocity of the other $n=5$
flow is $\vec{v}_5^+$, or vice versa.  In any case, neither $\vec{v}_5^-$
nor $\vec{v}_5^+$ lie close to the ecliptic plane, and hence the Earth
never approaches the spike caustic of the big flow.

Fig. 4 shows the speeds on Earth of the daughters of the big flow 
as a function of time of year if the big flow initial velocity is
$\vec{v}_5^-$.  The Earth enters the skirt near March 30 and exits 
it near Dec. 15.  So flows 2 and 3 only exist between those two
dates.  They are at any rate very hard to observe in a WIMP detector 
since their densities are suppressed by five orders of magnitude 
relative to flow 1.  Flow 1 has speed modulation of order 7\%.  The 
speed, and hence the flux on Earth, is largest near Nov. 10 and lowest
near May 10.

Fig. 5 shows the annual speed modulation of the big flow if its 
velocity is $\vec{v}_5^+$.  In this case also, flow 1 has a 
speed modulation of order 7\% but the maximum occurs near 
Jan. 20 and the minimum near July 20.

Table II shows two flows whose initial velocity vectors $\vec{v}_0$ 
are close to the ecliptic plane, namely 4+ and 10+.  For reasons mentioned
in Appendix B, the uncertainties on the flow velocities $\vec{v}^{n\pm}$
are larger for $n \leq 4$ than for $n \geq 5$. So, flow 4+ is less certain
to be close to the ecliptic plane than flow 10+ and therefore less certain
to show spiky behavior.  Figs. 6 and 7 show the velocity and density
modulation of flow 10+.  The behaviour near the spike is, of course,
sensitive to the precise value of $v_Z^{10+}$.  As an example, we chose
$v_Z^{10+} = 2.6$ km/s in Figs. 6 and 7.  The Earth approaches the spike
near Dec. 20.  The densities of flows 1 and 2 have a large temporary
enhancement ($F_{\rm max} \simeq 16$), and their velocities change
abruptly then.

The WIMP signal modulation for the caustic ring halo model is 
dominated by the speed modulation of the big flow.  This is 
evident in Figs. \ref{bf5-} and \ref{bf5+}, which show $T(E_r)$ 
at those times when the big flow speed is largest and smallest.  
In Fig. \ref{bf5-} the densities $d_n^\pm$ with  $n=5,6,7,8,9$ 
are all assigned the velocities $\vec{v}_n^\mp$, whereas in 
Fig. \ref{bf5+}, they are all assigned the velocities 
$\vec{v}_n^\pm$.  Each flow modulates as described in the 
previous subsection, with its individual phase and amplitude 
determined by its initial flow direction. 

To give a qualitative discussion let us neglect all but the big flow.  
To include the other flows is not difficult, but it complicates 
matters considerably without changing the results much.  The annual 
modulation depends sharply on the range, $E_1 \leq E_r \leq E_2$, of
measured recoil energies.  The edge of the big flow plateau as a 
function of time of year is 
\beq E_{\rm max}(t) = {2 m_r^2 \over m_N}~(v_5^\pm)^2~
\left[1 + {60~{\rm km/s} \over v_5^\pm} \sin\Theta_5^\pm~ 
\sin \left({2 \pi \over {\rm year}}(t - t_\pm)\right)\right] 
\label{Emt}
\eeq 
with either $v_5^- = 255$ km/s, $\Theta_5^- = 41^\circ$ and $t_- = t$ 
on Aug. 10, or $v_5^+ = 265$ km/s, $\Theta_5^+ = 36^\circ$ and $t_+ = t$ 
on Oct. 20, depending on whether the initial velocity of the big flow
is $\vec{v}_5^-$ or $\vec{v}_5^+$.  The height of the big flow plateau 
modulates with $\sim 7\%$ amplitude whereas the edge modulates with 
$\sim 15\%$ amplitude.  The plateau height and edge modulations have
opposite phase.  See Eqs. (\ref{diffrate}-\ref{Emax}). 

If $E_1$ and $E_2$ are both less than the minimum value $E_{\rm max,min}$
of $E_{\rm max}(t)$ in the course of the year, the observed event rate
modulates with amplitude of order 7\%, and maximum near either May 10 or
July 20.  This prediction of the caustic ring halo model is independent 
of the nuclear form factors involved. Such a modulation is very similar
to that which occurs in the isothermal halo model, which predicts 7\%
modulation with the maximum occurring near June 2.  This is so in spite 
of the fact that the isothermal and caustic ring models could hardly be
more dissimilar.  

For other values of $E_1$ and $E_2$, however, the rate modulation in the
caustic ring halo model is very different.  If $E_1$ and $E_2$ are both
larger than the maximum $E_{\rm max,max}$ of $E_{\rm max}(t)$ in the
course of year, the WIMP rate vanishes.  If $E_1$ is less than 
$E_{\rm max, min}$ and $E_2$ is larger than $E_{\rm max,max}$, the annual
modulation amplitude is 7\% (if $E_1 = 0$) or larger, with the
maximum occurring near November 10 or January 20.  If $E_1$ and/or $E_2$
fall between $E_{\rm max,min}$ and $E_{\rm max,max}$, the time derivative
of the WIMP event rate changes abruptly when $E_{\rm max}(t)$ equals $E_1$
or $E_2$.  In particular, if $E_{\rm max,min} < E_1 < E_{\rm max,max}$,
the signal is on part of the year and off for the remainder.  The
different kinds of behavior are easy to derive by combining Eqs.
(\ref{diff2}), (\ref{Emax}) and (\ref{Emt}).

The most dramatic prediction of the caustic ring model is that, 
for certain values of the recoil energy $E_r$, those between 
$E_{\rm max,min}$ and $E_{\rm max,max}$, ${dR \over dE_r}$
from the big flow (of order 75\% of the total local density)
is zero for a definite period of the year, and non-zero for the 
remainder.

A question of particular interest is whether the caustic ring halo
model is consistent with the annual modulation observed by the DAMA
experiment~\cite{DAMA}.  DAMA searches for WIMP dark matter by 
measuring the scintillation produced in NaI crystals when WIMPs 
scatter off nuclei of the crystal.  Of order 30\% (``quenching
factor" $q_{Na}$ = 0.3) of the nuclear recoil energy is measured as
scintillation if the WIMP scatters off a Na nucleus, whereas of order 
9\% is measured if the WIMP scatters off a I nucleus ($q_I$ = 0.09).  
Although Na has the larger quenching factor, in most particle physics
models the WIMP elastic scattering rate in NaI is dominated by 
scattering off I nuclei because the latter are heavier ($A_I$ = 127
vs. $A_{Na}$ = 23).  

DAMA observes an annual modulation of the scintillation rate with
amplitude of order 0.02 counts/day~kg~keV$_{ee}$ in the energy range
2-6 keV$_{ee}$.  The $ee$ subscript, which means 'electron equivalent', 
indicates that the scintillation energy is being given, not the nuclear 
recoil energy.  The modulation is consistent with being sinusoidal, with 
phase such that the maximum rate occurs on May 21 $\pm$ 22 days.  No 
significant annual modulation is observed in the 6-14 keV$_{ee}$ energy 
region.  

The annual modulation observed by DAMA in the 2-6 keV$_{ee}$
region is consistent with the caustic ring halo model provided 
$E_{\rm max}$ is large enough.  If the WIMP-nucleus interaction 
is spin-independent, the signal is dominated by scattering off I 
nuclei.  The requirement that $q_I~E_{\rm max} > 6$ keV implies  
that the WIMP mass is larger than about 190 GeV.  WIMP masses 
as small as 100 GeV may be accommodated depending on the details 
of cross-sections, quenching factors, and nuclear form factors.
Fig.~\ref{fig:anmod} shows the event rate modulations in the range
2-6 keV$_{ee}$ for the caustic ring halo model with the big flow
velocity equal to $\vec{v}_5^-$ for $m_\chi=120$ GeV and 
$m_\chi=50$ GeV, assuming the WIMP-nucleus interaction is 
spin-independent and the nuclear form factors are equal to one.

The lack of annual modulation observed by DAMA in the 6-14 keV$_{ee}$
range, given that DAMA does observe a modulation in the 2-6 keV$_{ee}$
range, may be hard to accommodate in the caustic ring model.  It would 
require some degree of cancellation among competing effects, e.g. a 
modulation with one phase during part of the year and with the opposite 
phase during the remainder.  This can occur if $q_I~E_{\rm max}$ is 
close to 14 keV.  However to compare the model with the DAMA data in 
such scenarios one would have to know the details of the nuclear form 
factors and make some assumptions about the particle physics.  Such a 
comparison is beyond the scope of this paper. 

Instead let us emphasize that DAMA can test the caustic ring halo 
model by measuring ${dR \over dE_r}(E_r, t)$ and asking whether for 
some values of $E_r$, of order 75\% of the event rate is off part of 
the year and on for the remainder, as predicted by Eq. (\ref{Emt}).

\section{Conclusions}

For an arbitrary single cold flow, dubbed a 'mother flow', of initial
velocity $\vec{v}_0$ and density $d_0$ far from the Sun, we derived the
densities $d_i(t)$ and speeds $v_i(t)$ on Earth of the daughter flows as a
function of time of day and time of year.  We included all corrections
down to the $10^{-3}$ level of accuracy.  In order of decreasing
importance, the following effects are relevant: the orbital motion of the
Earth (10\%), the effect of solar gravity (typically 1\% but of order
100\% or even larger if the Earth approaches a caustic), the Earth's
rotation (0.3\%), and the Earth's gravity (0.1\%).  Although we did not
systematically study the next set of corrections, it seems that they are
all much smaller.  We expect the asphericity of the Earth and that of the
Sun to produce corrections of order $10^{-5}$, and we already mentioned
that Jupiter introduces corrections of order $10^{-6}$.  So, it may be
that our calculations are in fact accurate at the $10^{-5}$ level of
precision.
 
Our results are directly applicable to discrete dark matter flows and
streams.  However, they can also be applied to an arbitrary, continuous
and/or discrete, dark matter velocity distribution $f(\vec{v}_0)$, 
simply by integrating over $\vec{v}_0$.  

The flows predicted by the caustic ring halo model were discussed as 
an example.  They are listed in Tables I and II.  The model predicts 
that the local velocity distribution is dominated by a single 
flow, dubbed the 'big flow'.   It is one of the flows associated 
with the caustic ring of dark matter nearest to us.  There is a two-fold 
ambiguity in the velocity of the big flow.  In one case ($\vec{v}^{5-}$),
the big flow initial velocity has magnitude 255 km/s, and is pointing in
the direction of declination $\delta_5^- = 31^\circ$ and right ascension
$\alpha_5^- = 20.2^{\rm h}$.  In the second case ($\vec{v}^{5+}$), its 
initial velocity has magnitude 265 km/s, and direction $\delta_5^+ = 
60^\circ,~\alpha_5^+ = 23.8^{\rm h}$.

In the cavity detector of dark matter axions a cold flow produces a 
narrow peak in the spectrum of microwave photons from axion conversion.  
A peak's frequency can readily be measured with an accuracy of $10^{-11}$.  
This implies an accuracy of order $10^{-5}$ in the flow speeds.  [There 
is no difficulty in measuring a peak's frequency with $10^{-14}$ accuracy 
using atomic clocks.  However, in the 100 sec signal integration time 
typical of the experiment in search mode, a peak shifts by an amount 
of order $10^{-11}$; see Eq. (\ref{rs})].  The existence of the big flow
allows the axion experiment to be made more sensitive by searching for 
the corresponding narrow peak.  This capability has already been 
implemented in the ADMX experiment \cite {ADMX}. If a signal is 
found, it will be possible to measure for each mother flow the 
densities and speeds of the three daughter flows year round.  Since 
the properties of flows 2 and 3 near the skirt caustic depend on the 
mass distribution inside the Sun, the latter may be investigated.

In a WIMP detector, a cold flow produces a plateau in the recoil 
energy spectrum of nuclei off of which WIMPs have scattered elastically.  
The edge of this plateau is the maximum recoil energy for given flow 
speed [see Eq. (\ref{Emax})]; it is proportional to the flow speed
squared.  On the other hand, the height of the plateau is proportional 
to the (flow speed)$^{-1}$.  The annual modulation due to a single
cold WIMP flow depends on the range of observed recoil energies.  If 
the whole observed range is below the plateau edge, the event rate
modulation has the opposite phase and the same relative amplitude as 
the speed modulation.  If the whole plateau is in the observed range, 
the event rate modulation has the same phase and the same relative 
amplitude as the speed modulation.  If only part of the plateau is 
in the observed range, the behaviour is more complicated.  The speed
modulation of the big flow has amplitude of order 7\%. The maximum 
speed occurs near May 20 if the big flow velocity is $\vec{v}^{5-}$, 
July 20 if the big flow velocity is $\vec{v}^{5+}$.

The most dramatic prediction of the caustic ring model for WIMP 
searches is that, for a certain range of recoil energies, the 
event rate due to the big flow (of order 75\% of the signal)
is on during a predicted part of the year and off for the remainder.

The DAMA results published so far are not inconsistent with the caustic
ring halo model.  The modulation observed in the 2-6 keV$_{\rm ee}$ 
range is consistent with the model provided the WIMP mass is large 
enough ($\gsim 100$ GeV).  The lack of observed modulation in the 
6-14 keV$_{\rm ee}$ range may be inconsistent with the model, 
but to actually show this one needs to know the nuclear form 
factors and to make some assumptions about the particle physics
model. Instead, DAMA may simply test the caustic ring halo model 
by looking for the predicted big flow plateau in the recoil energy
spectrum and the predicted annual modulation of that plateau.

\section{Acknowledgments}

P.S. would like to thank Laura Baudis, Leanne Duffy and the members 
of the ADMX collaboration for illuminating discussions, and the Aspen 
Center for Physics for its hospitality while working on this project.  
The work of F-S.L.~and P.S.~was supported in part by the U.S. 
Department of Energy under grant DE-FG02-97ER41029.  The work of 
S.W. at NRL was supported by the NASA GLAST Science Investigation 
Grant DPR S-1536-Y.  The work of F.-S. L. in Brussels was supported 
by the I.I.S.N. and the Belgian Science Policy (return grant and IAP
5/27).


\appendix

\section{Coordinate systems}

Here we enumerate the different coordinate systems used in the 
main body of the paper and state how they are related to one 
another.  We use four coordinate systems: galactic, solar,
terrestrial and "wake".  All four coordinate systems are 
right-handed.

The galactic coordinates are $(x_G, y_G, z_G)$ with $\hat{x}_G$ 
in the local direction away from the galactic center, $\hat{y}_G$ 
in the direction of galactic rotation, and $\hat{z}_G$ in the 
direction of the South Galactic pole.  

The solar coordinates are $(X, Y, Z)$ with $\hat{Z}$ perpendicular 
to the plane of the Earth's orbit ("the ecliptic"), $\hat{X}$ in the
direction of the Earth relative to the Sun at the time of the autumnal
equinox (near Sept. 23), and $\hat{Y}= \hat{Z} \times \hat{X}$.  The
corresponding polar coordinates are called $(\Theta, \Phi)$.

The terrestrial coordinates are $(x_\oplus, y_\oplus, z_\oplus)$, 
where $\hat{z}_\oplus$ is the Earth's rotation
axis, $\hat{x}_\oplus$ is in the same direction as $\hat{X}$,
and $\hat{y}_\oplus = \hat{z}_\oplus \times \hat{x}_\oplus$.
The corresponding polar coordinates $(\theta_\oplus, \phi_\oplus)$
are related to right ascension $\alpha$ and declination 
$\delta$ by $\theta_\oplus = {\pi \over 2} - \delta$ and 
$\phi_\oplus = \alpha$.

The relation between terrestrial and solar coordinates follows from
the fact that the ecliptic is inclined at $\gamma = 23.44^\circ$
relative to the Earth's equator.  We have
\beq
\left( \ba{c}
X \\ Y \\ Z \ea
\right) =
\left(
\ba{ccc}
1 & 0 & 0\\
0 & \cos \gamma & \sin \gamma \\
0 & -\sin \gamma & \cos \gamma
\ea
\right)
\left(
\ba{c}
x_\oplus \\ y_\oplus \\ z_\oplus \ea
\right)~~~\ .
\eeq
The relation between solar and galactic coordinates follows 
from the right ascension and declination of the Galactic
Center (GC) and Galactic North Pole (GNP) \cite{eq00}: 
$\alpha_{\rm GC} = 17 {\rm h}~45.6 {\rm m}~,~
\delta_{\rm GC} = - 28^\circ~56.2'~,~\alpha_{\rm GNP} = 
12 {\rm h}~51 {\rm m}~,~\delta_{\rm GNP} = 27^\circ~07.7'$.  
This yields
\beq
\left( \ba {c} X \\ Y \\ Z \ea \right)
=\left(\ba {ccc}.055 & .494 & .868 \\
.994 & -.111 & .000 \\
.096 & .862 & -.497 \ea\right)
\left( \ba {c} x_G \\ y_G \\ z_G
\ea\right).
\label{eq:transform}
\eeq

To describe the effect of Sun's gravity on a particular dark matter 
flow, it is convenient to use "wake" coordinates.  These are cylindrical 
coordinates $(\rho, \phi, z)$, with $\hat{z}$ axis is the direction 
of the flow velocity $\vec{v}_0$ far upstream of the Sun. Unlike 
ordinary cylindrical coordinates, however, we let $\rho$
range from $-\infty$ to $+\infty$, and $\phi$ from 0 to $\pi$.
This avoids the sudden change in $\phi$ value of the dark matter 
particles when they cross the $z$-axis.

\section{The caustic ring halo model}

The velocities and densities of the first 40 cold dark matter flows 
predicted by the caustic ring halo model at the Earth's location are 
listed in Table I.  The velocities are with respect to the non-rotating 
rest frame of the Galaxy, and are given in the galactic coordinate 
system defined in Appendix A.  Table I is an update of the table 
published in ref. \cite{bux}.  The main difference is that the 
fifth flow is much more prominent in the new version.

The caustic ring model explicitly realizes the fact that all 
cold dark matter particles lie on a three-dimensional sheet in 
six-dimensional phase-space.  As a corollary, there is a discrete 
set of flows at any location in a galactic halo.  The number of 
flows is location dependent.  There are caustic surfaces wherever 
the local number of flows changes by two.  At the caustics the 
density of dark matter is very large.  It was shown that two types of 
caustics, called {\it inner} and {\it outer}, are necessarily present 
in every galactic halo \cite{PS2}.  The outer caustics are simple 
fold ($A_2$) catastrophes located on topological spheres surrounding 
the galaxy.  The inner caustics are elliptic umbilic ($D_{-4}$) 
catastrophes located on closed tubes lying in the galactic
plane.  The inner caustics are usually referred to as "caustic 
rings of dark matter".  One caustic ring is associated with each 
flow in and out of the Galaxy. 

Observational evidence for caustic rings has been found in the 
distribution of rises in the Galactic rotation curve and in the 
presence of a triangular feature in the IRAS map of the Galactic 
plane \cite{mw}.  The list of flows in Table I results from a fit 
to these observations. We do not justify Table I here, but merely 
use its flows as examples.

The flows are labeled by an integer $n=1,2,3,..$ and a binary symbol
$\pm$.  In the caustic ring halo model, the first eight flows at the
Earth's location $(n = 1,2,3,4)$ have large velocity components in the
$\pm \hat{x}_G$ and $+ \hat{y}_G$ directions, whereas the remaining flows
$(n \geq 5)$ have large components in the $\pm \hat{z}_G$ and $+
\hat{y}_G$ directions.  The two $n=5$ flows have the largest local
densities, by far.  The reason is that the $n=5$ caustic ring of dark
matter, whose cross-section is allegedly revealed in the aforementioned
IRAS map, is very close to us.  One of the $n=5$ flows has of order three
times as much local density as all the other flows combined.  So, it was
named the "big flow".  The density of the other $n=5$ flow is much smaller
than that of the big flow, but much larger than that of any of the other
$n \neq 5$ flows.  The density of the big flow is large because of our
proximity to a cusp in the nearby caustic ring.  It is sensitive to our
distance to that cusp.  Because that distance is poorly known, there is a
large uncertainty on the density of the big flow.  Nonetheless it is clear
that the big flow dominates over all other flows combined.
 
For $n \leq 4$, $d_n^+ = d_n^-$ because the model is reflection 
symmetric.  For $n \geq 5$, $d_n^+$ and $d_n^-$ are different in 
general.  For $n = 5$ through 9, different estimates are given for 
$d_n^+$ and $d_n^-$, but for $n \geq 10$ the differences between 
$d_n^+$ and $d_n^-$ are not deemed sufficiently important to 
differentiate them.  At present it is not possible to say, for each
$n \geq 5$, whether $d_n^\pm$ is the density of the flow of velocity
$\vec{v}^{n \pm}$ or $\vec{v}^{n \mp}$.  In particular this is so for
$n=5$.  The velocities of the $n=5$ flows are
\beq
\vec{v}^{5\pm}=(470~\hat{y}_G \pm 100~\hat{x}_G)\rm{km/s}
\label{v5}
\eeq
in galactic coordinates.  However it is not known whether $\vec{v}_5^-$
or $\vec{v}_5^+$ is the velocity of the big flow.

For $n \leq 4$, no entries are given for $v^{n\pm}_{xG}$ because 
not enough information is presently available to estimate those 
velocity components.  However the $v^{n\pm}_{xG}$ velocity 
components are expected to be relatively small, $\lesssim 30\%$ 
of the corresponding velocity magnitudes $v^{n\pm}_G$.

In the rest frame of the Sun, the flow velocities are 
\beq 
\vec{v}^{n\pm}_\odot = \vec{v}^{n\pm} - \vec{v}_{\odot} 
\label{eq:4.2} 
\eeq 
where 
\beq 
\vec{v}_{\odot}=(-9 \hat{x}_G +232 \hat{y}_G - 
7 \hat{z}_G)\rm{km/sec}
\label{eq:4.3} 
\eeq 
is the velocity of the Sun with respect to the Galaxy.  In
Eq.~(\ref{eq:4.3}) we assume that the Sun moves at 220 km/sec  
in the direction $\hat{y}_G$ of galactic rotation plus
16.5 km/sec in the direction of galactic longitude $l=53^{\circ}$
and latitude $b=25^{\circ}$ \cite{BT}.  The velocity components 
of the flows in solar coordinates, obtained from Table I by 
applying Eqs. (\ref{eq:4.2}) and (\ref{eq:transform}), are
listed in Table II.  For the sake of definiteness we have set
$v^{n\pm}_{xG} = 0$ when computing the $n= 1 .. 4$ entries in 
Table II.

 
\newpage      
\begin{table}
\caption{Velocity vectors $\vec{v}^{n\pm}$ and densities $d_n^\pm$
of the first 40 flows in the caustic ring halo model, in galactic 
coordinates. The flow of velocity vector $\vec{v}^{n\pm}$ has 
density $d_n^\pm$ or $d_n^\mp$.}
\begin{center}
\footnotesize  
\begin{tabular}{|r|c|c|c|c|c|c|} 
\hline
$n$&$v^{n\pm}_G$&$v^{n\pm}_{yG}$&$v^{n\pm}_{zG}$&$v^{n\pm}_{xG}$&
$d_n^{+}$&$d_n^{-}$\\
&(km/s)&(km/s)&(km/s)&(km/s)&($10^{-26}$gr/cm$^3$)&($10^{-26}$gr/cm$^3$)\\
\hline
1 & 620& 130 & $\pm$605 & ~~/~~ & 0.3 & 0.3 \\
2 & 560 & 230 & $\pm$510 & ~~/~~ & 0.8 & 0.8 \\ 
3 & 530 & 320 & $\pm$420 & ~~/~~ & 1.4 & 1.4 \\
4 & 500 & 405 & $\pm$300 & ~~/~~ & 3.4 & 3.4\\
5 & 480 & 470 & ~~0~~ & $\pm$100 & 170. & 15. \\   
6 & 465 & 400 & ~~0~~ & $\pm$240 & 6.5 & 3.4 \\
7 & 450 & 330 & ~~0~~ & $\pm$305 & 4.1 & 1.3 \\
8 & 430 & 295 & ~~0~~ & $\pm$320 & 2.0 & 1.1 \\ 
9 & 420 & 240 & ~~0~~ & $\pm$340 & 1.5 & 0.7 \\
10 & 410 & 200 & ~~0~~& $\pm$355 & 1.0 & 1.0 \\
11 & 395 & 180 & ~~0~~& $\pm$350 & 0.9 & 0.9 \\
12 & 385 & 160 & ~~0~~& $\pm$350 & 0.8 & 0.8 \\
13 & 375 & 150 & ~~0~~& $\pm$345 & 0.7 & 0.7 \\ 
14 & 365 & 135 & ~~0~~& $\pm$340 & 0.7 & 0.7 \\
15 & 355 & 120 & ~~0~~& $\pm$335 & 0.6 & 0.6 \\
16 & 350 & 110 & ~~0~~& $\pm$330 & 0.6 & 0.6 \\
17 & 340 & 105 & ~~0~~& $\pm$320 & 0.5 & 0.5 \\
18 & 330 & ~95 & ~~0~~& $\pm$315 & 0.5 & 0.5 \\ 
19 & 320 & ~90 & ~~0~~& $\pm$310 & 0.5 & 0.5 \\
20 & 310 & ~80 & ~~0~~& $\pm$300 & 0.4 & 0.4 \\
\hline 
\end{tabular} 
\vspace{0.2cm} 
\end{center}
\label{t:table1}
\end{table} 
 
\begin{table}[t] 
\caption{Densities and velocity vectors of the first 40 flows 
in the caustic ring halo model, in solar coordinates.}
\begin{center} 
\footnotesize 
\begin{tabular}{|r|c|c|c|c|c|} 
\hline 
$n\pm$&$v^{n\pm}_\odot$&$v^{n\pm}_X$&$v^{n\pm}_Y$&$v^{n\pm}_Z$&$d_n^\pm$\\
&(km/s)&(km/s)&(km/s)&(km/s)&($10^{-26}$gr/cm$^3$) \\
\hline 
1+  & 620 & 480  & ~20  & -395 & 0.3 \\ 
1-- & 605 & -570 & ~20  & 210  & 0.3 \\ 
2+  & 520 & 450  & ~10  & -255 & 0.8 \\ 
2-- & 505 & -440 & ~10  & 250  & 0.8 \\ 
3+  & 435 & 415  & $\sim$ 0 & -130. & 1.4 \\ 
3-- & 420 & -310 & $\sim$ 0 & 285 & 1.4 \\ 
4+  & 350 & 350  & -10  & $\sim$ 0 & 3.4 \\ 
4-- & 340 & -165 & -10  & 295  & 3.4 \\ 
5+  & 265 & 130  & ~85  & 210  & 15. {\rm or} 170.\\ 
5-- & 255 & 120  & -120 & 195  & 170. {\rm or} 15.\\ 
6+  & 300 & 100  & 230  & 160  & 3.4 {\rm or} 6.5\\ 
6-- & 280 & 75   & -245 & 120  & 6.5 {\rm or} 3.4 \\ 
7+  & 330 & 70   & 300  & 110  & 1.3 {\rm or} 4.1 \\
7-- & 310 & 35   & -305 & 50   & 4.1 {\rm or} 1.3 \\
8+  & 330 & 55   & 320  & 80   & 1.1 {\rm or} 2.0 \\
8-- & 315 & 20   & -310 & 20   & 2.0 {\rm or} 1.1 \\
9+  & 350 & 30   & 350  & 40   & 0.7 {\rm or} 1.5 \\
9-- & 330 & -10  & -330 & -25  & 1.5 {\rm or} 0.7 \\
10+ & 365 & 10   & 365  & $\sim$ 0 & 1.0 \\
10--& 350 & -30  & -340 & -60  & 1.0 \\
11+ & 365 & $\sim$ 0 & 360 & -15 & 0.9 \\ 
11--& 345 & -40  & -335 & -80  & 0.9 \\ 
12+ & 365 & -10  & 360  & -30  & 0.8 \\
12--& 345 & -50  & -330 & -95  & 0.8 \\
13+ & 360 & -15  & 360  & -40  & 0.7 \\
13--& 345 & -55  & -325 & -105 & 0.7 \\
14+ & 360 & -20  & 355  & -55  & 0.7 \\
14--& 345 & -60  & -320 & -120 & 0.7 \\ 
15+ & 360 & -30  & 350  & -65  & 0.6 \\ 
15--& 340 & -65  & -310 & -130 & 0.6 \\
16+ & 360 & -35  & 350  & -75  & 0.6 \\
16--& 340 & -70  & -305 & -135 & 0.6 \\
17+ & 355 & -40  & 340  & -80  & 0.5 \\
17--& 340 & -75  & -300 & -140 & 0.5 \\
18+ & 350 & -40  & 340  & -90  & 0.5 \\
18--& 335 & -80  & -290 & -150 & 0.5 \\
19+ & 350 & -50  & 330  & -100 & 0.5 \\
19--& 330 & -80  & -280 & -155 & 0.5 \\
20+ & 345 & -50  & 325  & -105 & 0.4 \\
20--& 330 & -85  & -275 & -160 & 0.4 \\
\hline 
\end{tabular} 
\end{center} 
\label{t:table2} 
\end{table} 
 


\begin{figure}
\epsfxsize=4in
\centerline{\epsfbox{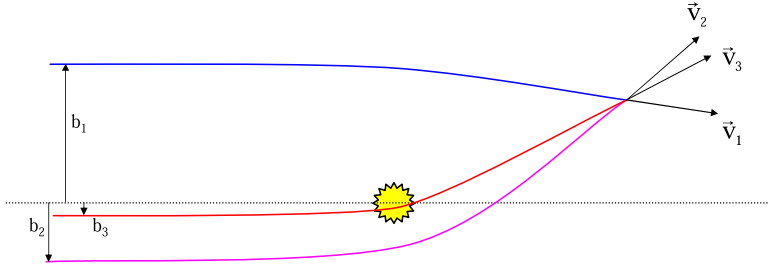}}
\vspace{1.cm}
\caption{Schematic illustration of the effect of solar gravity 
on a CDM flow.  Downstream of the Sun there are three flows, 
whereas upstream there is only one flow.  The surface separating 
the two regions is the location of a caustic called the 'skirt'.
It is a cone with opening angle equal to the maximum scattering 
angle of particles in the flow.}
\label{fig:suncdm}
\end{figure}

\begin{figure}
\epsfxsize=4in
\centerline{\epsfbox{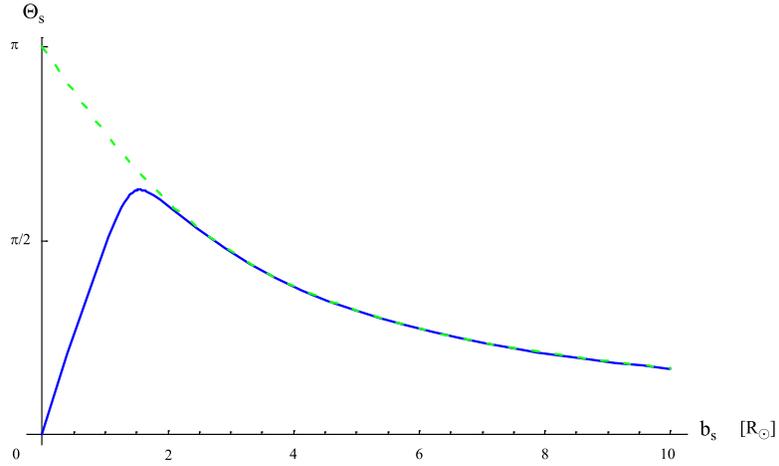}}
\vspace{1.cm}
\caption{The scattering angle $\Theta_{\rm s}$ versus impact
parameter $b_{\rm s}$, for the solar model of ref. [30] and 
incoming flow speed $v_0 = 265$ km/s.  The dotted line 
is for a point mass Sun.}
\label{fig:Thvsbs}
\end{figure}

\begin{figure}
\epsfxsize=4in
\centerline{\epsfbox{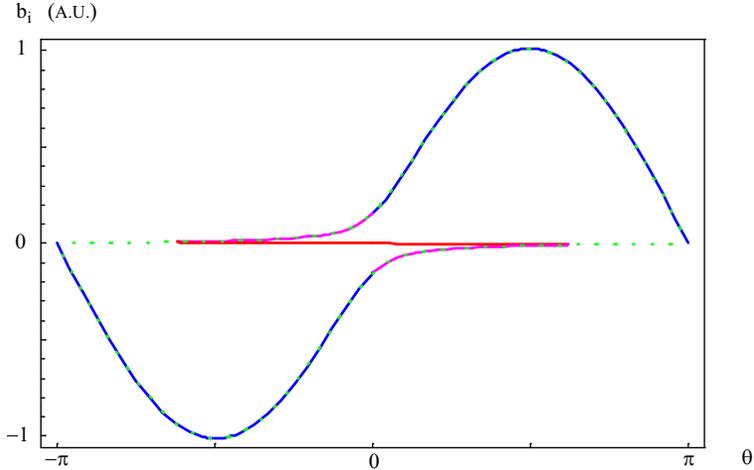}}
\vspace{1.cm}
\caption{The impact parameters $b_i~(i=1,2,3)$ versus polar angle, 
defined relative to the incoming flow direction $\hat{v}_0$, at the 
Earth's distance from the Sun, for the solar model of ref. [30] and 
incoming flow speed $v_0 = 265$ km/s.  The dotted line is for a 
point mass Sun.  When $|\theta|$ exceeds the maximum scattering 
angle, there is only one flow and hence only one value ($b_1$) of 
the impact parameter.  When $|\theta|$ is less than the maximum 
scattering angle, there are three flows and three values ($b_1, b_2$ 
and $b_3$) of the impact parameter.}
\label{fig:dvsth}
\end{figure}

\begin{figure}
\epsfxsize=4in
\centerline{\epsfbox{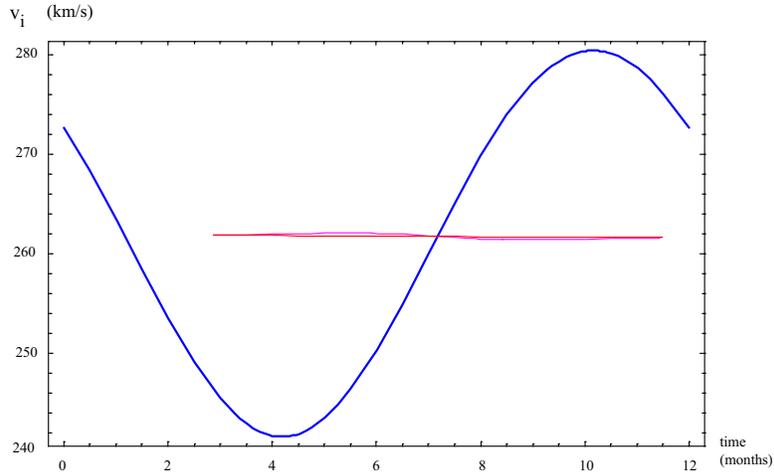}}
\vspace{1.cm}
\caption{Flow speeds measured on Earth if the initial flow velocity 
is $\vec{v}^{5-}$, as a function of time of year, since Jan. 1.  We 
neglect here the Earth's rotation and the eccentricity of its orbit.
Flows 2 and 3 exist on Earth from approximately March 30 until
approximately Dec. 15.  Their speed modulation is very small.  
Flow 1 exists throughout the year.  Its speed modulation is of 
order 7\%.  The speed on Earth of flow 1 is largest near November 10 
and lowest near May 10.}
\label{fig:v5-}
\end{figure}

\begin{figure}
\epsfxsize=4in
\centerline{\epsfbox{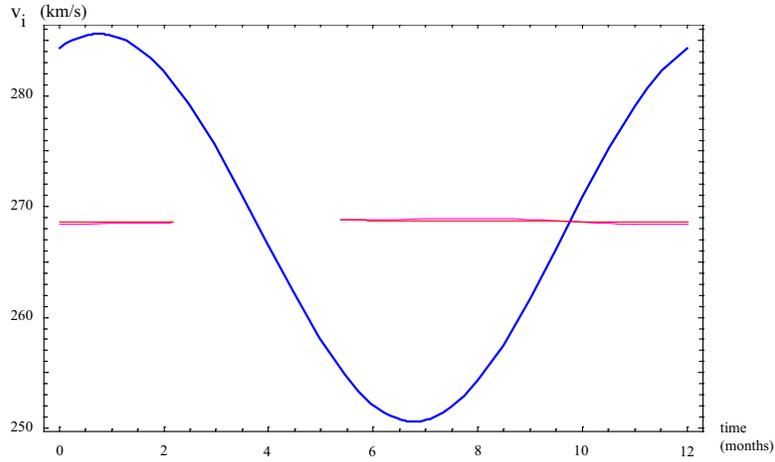}}
\vspace{1.cm}
\caption{Same as Fig. \ref{fig:v5-} but with initial flow velocity 
equal to $\vec{v}^{5+}$.  Flows 2 and 3 exist on Earth from 
approximately June 13 until approximately March 7.  The speed 
of flow 1 is largest near Jan. 20 and lowest near July 20.} 
\label{fig:v5+}
\end{figure}

\begin{figure}
\epsfxsize=4in
\centerline{\epsfbox{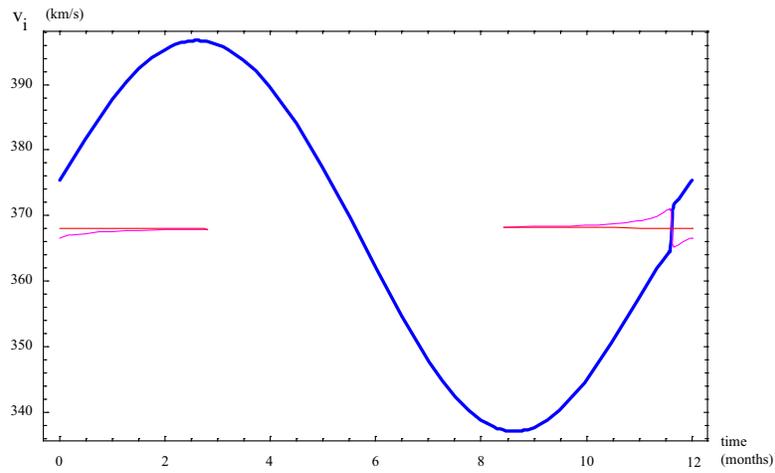}}
\vspace{1.cm}
\caption{Same as Figs. \ref{fig:v5-} and \ref{fig:v5+} but with 
initial flow velocity equal to $\vec{v}^{10+}$, whose direction 
is close to the ecliptic.  The Earth passes near the flow's spike 
caustic on approximately Dec. 20.}  
\label{fig:v10+}
\end{figure}

\begin{figure}
\epsfxsize=4in
\centerline{\epsfbox{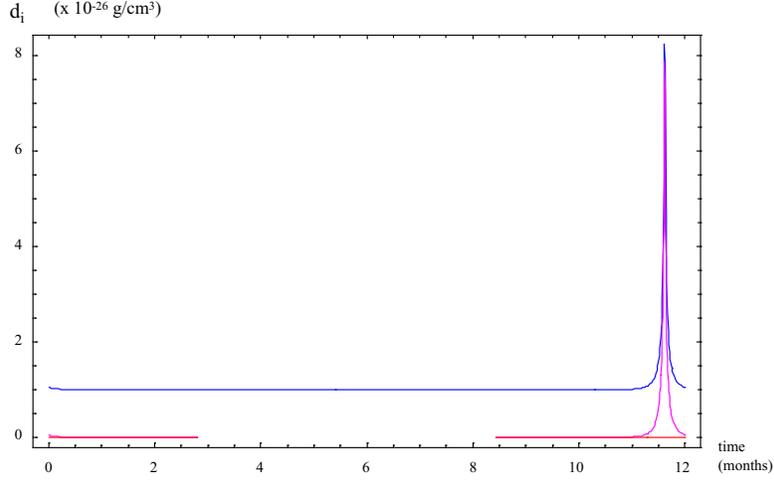}}
\vspace{1.cm}
\caption{Flow densities $d_i~(i=1,2,3)$ on Earth for flow 10+, 
as a function of time of year.  The flow speeds for this case
are shown in Fig. \ref{fig:v10+}.  The Earth passes near the 
flow's spike caustic on approximately Dec. 20.  If $v_Z^{10+} = 0$ 
the densities of flows 1 and 2 diverge at that moment.  In 
Figs. \ref{fig:v10+} and \ref{fig:d10+}, we used 
$v_Z^{10+} = 2.6$ km/s, as an example.  The Earth passes 
through the skirt caustic on approximately September 15 
and March 25.  However the associated divergences of the
density of flows 2 and 3 are too feeble to show up on 
the graph.}
\label{fig:d10+}
\end{figure}

\begin{figure}
\epsfxsize=4in
\centerline{\epsfbox{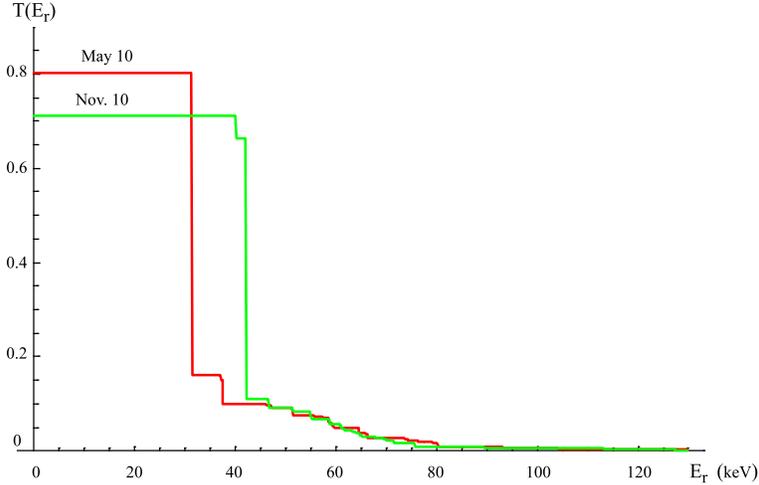}}
\vspace{1.cm}
\caption{The function $T(E_r)$, defined in Eq. (\ref{TE}), for 
the caustic ring model if the big flow has velocity $\vec{v}^{5-}$,
at those times of year when the speed of the big flow relative to 
Earth is largest(May 10) and smallest (Nov. 10).  For all $n$, 
the density $d_n^\pm$ was taken to be that of the flow with velocity 
$\vec{v}^{n\mp}$.  To establish the scale on the horizontal axis, 
we set the WIMP mass $m_\chi = 100$ GeV and target nucleus mass 
$m_N = A m_{\rm proton}$ with $A = 73$.  For different values of 
$m_\chi$ and $A$, rescale the horizontal axis by a factor proportional 
to $m_r^2/m_N$ where $m_r$ is the reduced mass.}
\label{bf5-}
\end{figure}

\begin{figure}
\epsfxsize=4in
\centerline{\epsfbox{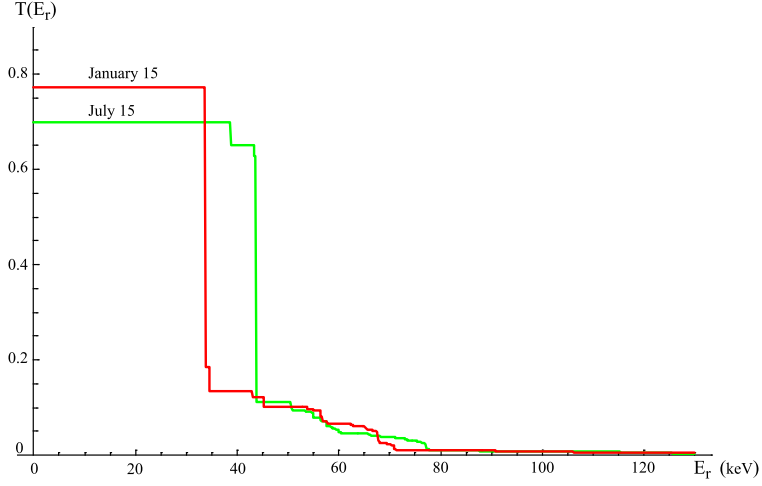}}
\vspace{1.cm}
\caption{Same as Fig. \ref{bf5-} except that now the big flow has 
velocity $\vec{v}^{5+}$ and, for all $n$, $d_n^\pm$ is taken to 
be the density of the flow with velocity $\vec{v}^{n\pm}$.}
\label{bf5+}
\end{figure}

\begin{figure}
\epsfxsize=4in
\centerline{\epsfbox{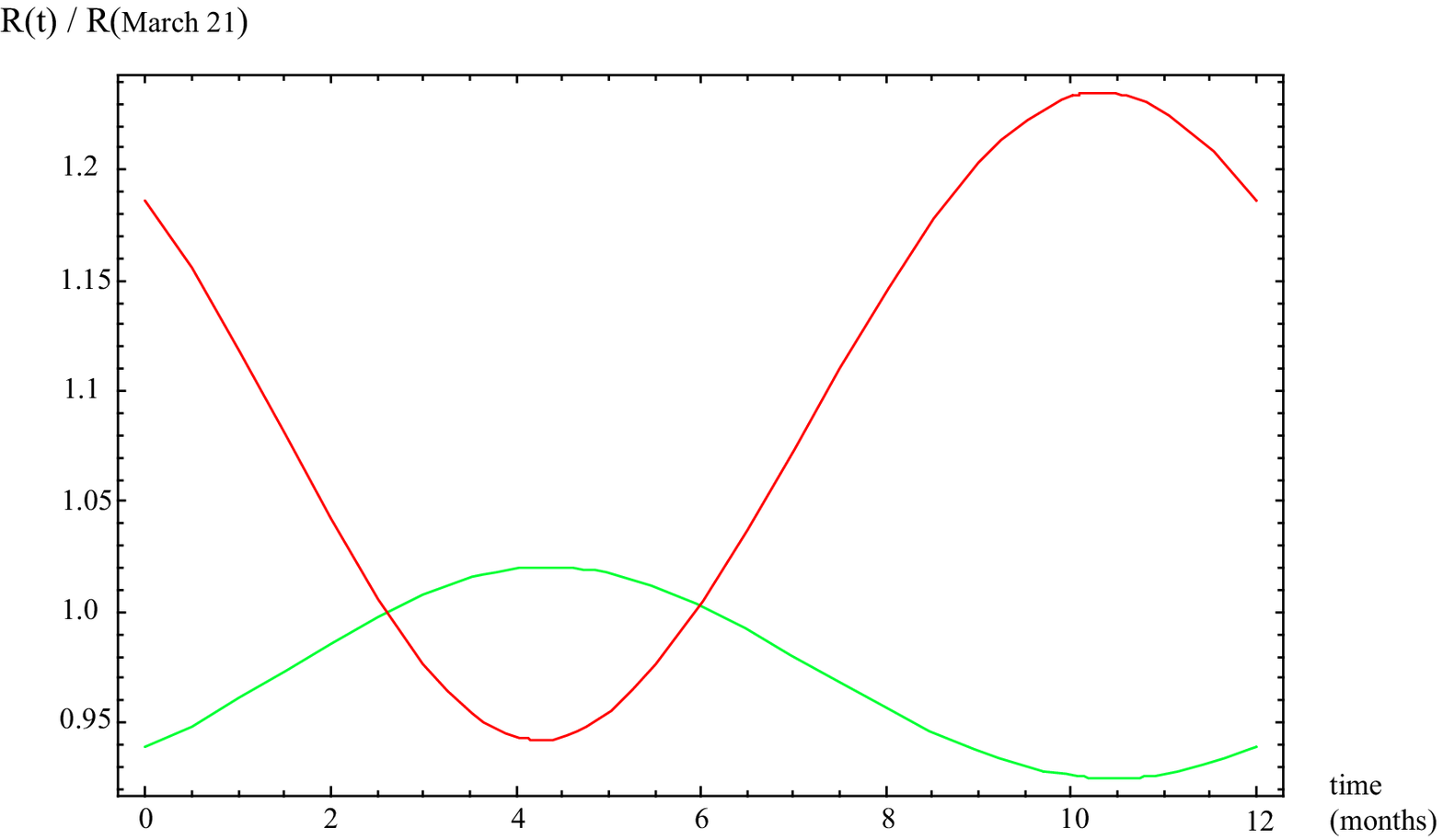}}
\vspace{1.cm}
\caption{Event rate for the DAMA experiment in the 2-6 keV$_{ee}$ 
range, as a function of time of year, for $m_\chi = 50$ GeV (top 
curve) and $m_\chi = 120$ GeV (bottom curve), relative to the 
event rate on March 21.  The velocity of the big flow was taken 
to be $\vec{v}^{5-}$, as in Fig. \ref{bf5-}.  Spin-independent
WIMP-nucleus couplings were assumed and the form factors were 
set equal to one.}  
\label{fig:anmod}
\end{figure}


\begin{references}

\bibitem{WMAP} C.L. Bennett et al., Ap. J. Suppl. 148 (2003) 1.

\bibitem{accel} S. Perlmutter et al., Ap. J. 517 (1999) 565;
A. Riess et al., Astron. J. 116 (1998) 1009. 

\bibitem{Zw} F. Zwicky, Helv. Phys. Act. 6 (1933) 110.

\bibitem{dgal} V.C. Rubin and W.K. Ford, Ap. J. 159 (1970) 379;
D.H. Rogstad and G.S. Shostak, Ap. J. 176 (1972) 315.

\bibitem{cold} P.J.E. Peebles, Ap. J. 263 (1982) L1;
J. Ipser and P. Sikivie, Phys. Rev. Lett. 50 (1983) 925;
G.R. Blumenthal, S.M. Faber, J.R. Primack and M.J. Rees, 
Nature 311 (1984) 517.

\bibitem{neutrino} S. Tremaine and J.E. Gunn, Phys. Rev. Lett. 42
(1979) 407; J.R. Bond, G. Efstathiou and J. Silk, Phys. Rev. Lett.
45 (1980) 1980.

\bibitem{axion} R.D. Peccei and H. Quinn, Phys. Rev. Lett. 38 (1977) 
1440 and Phys. Rev. D16 (1977) 1791; S. Weinberg, Phys. Rev. Lett. 40
(1978) 223; F. Wilczek, Phys. Rev. Lett. 40 (1978) 279.

\bibitem{axcosm} J. Preskill, M. Wise and F. Wilczek, Phys. Lett. 120B 
(1983) 127;  L. Abbott and P. Sikivie, Phys. Lett. 120B (1983) 133;
M. Dine and W. Fischler, Phys. Lett. 120B (1983) 137.  For a
discussion of the various contributions to the axion cosmological 
energy density, see S. Chang, C. Hagmann and P. Sikivie, Phys. Rev. 
D59 (1999) 023505. 

\bibitem{neutralino}
H. Goldberg, Phys. Rev. Lett. 50 (1983) 1419;
J. Ellis et al., Nucl. Phys. B238 (1984) 453.

\bibitem{WIMP} P. Hut, Phys. Lett. 69B (1977) 179;
B.W. Lee and S. Weinberg, Phys. Rev. Lett. 39 (1977) 183;
M.I. Vysotskii, A.D. Dolgov and Y.B. Zel'dovich, JETP Lett. 26 (1977) 165;
J.E. Gunn et al., Ap. J. 223 (1978) 1015.

\bibitem{axdet}
P. Sikivie, Phys. Rev. Lett. 51 (1983) 1415, and Phys. Rev. D32
(1985) 2988; S. DePanfilis et al., Phys. Rev. Lett. 59 (1987) 839;
C. Hagmann et al., Phys. Rev. D42 (1990) 1297;
I. Ogawa, S. Matsuki and K. Yamamoto, Phys. Rev. D53 (1996) 1740;
R. Bradley et al., Rev. Mod. Phys. 75 (2003) 777.

\bibitem{ADMX}
C. Hagmann et al., Phys. Rev. Lett. 80 (1998) 2043;
S. Asztalos et al., Phys. Rev. D64 (2001) 092003;
S. Asztalos et al., Ap. J. 571 (2002) L27. 

\bibitem{Wdet}
M.W. Goodman and E. Witten, Phys. Rev. D31 (1985) 3059;
I. Wasserman, Phys. Rev. D33 (1986) 2071;
A. Drukier, K. Freese and D.N. Spergel, Phys. Rev. D33 (1986) 3495.

\bibitem{Wrev}
J.R. Primack, D. Seckel and B. Sadoulet, Ann. Rev. Nucl. Part. Sci.
38 (1988) 751; P.F. Smith and J.D. Lewin, Phys. Rep. 187 (1990) 203;
G. Jungman, M. Kamionkowksi and K. Griest, Phys. Rep. 267 (1996) 195.

\bibitem{iso}
J.N. Bahcall and R.M. Soneira, Ap. J. Suppl. 44 (1980) 73;
J.A.R. Caldwell and J.P. Ostriker, Ap. J. 251, (1981) 61;
M.S. Turner, Phys. Rev. D33 (1986) 889;
R.A. Flores, Phys. Lett. B215 (1988) 73.

\bibitem{ann}
A. Drukier, K. Freese and D. Spergel, Phys. Rev. D33 (1986) 3495;
K. Freese, J. Frieman and A. Gould, Phys. Rev. D37 (1988) 3388.

\bibitem{vio}
D. Lynden-Bell, Mon. Not. R. Astron. Soc. 136 (1967) 101.

\bibitem{ips}
P. Sikivie and J.R. Ipser, Phys. Lett. B291 (1992) 288.

\bibitem{stw}
P. Sikivie, I. Tkachev and Y. Wang, Phys. Rev. Lett. 75 (1995) 2911,
and Phys. Rev. D56 (1997) 1863.

\bibitem{ss}
J.A. Filmore and P. Goldreich, Ap. J. 281 (1984) 1;
E. Bertschinger, Ap. J. Suppl. 58 (1985) 39.

\bibitem{PS1}
P. Sikivie, Phys.~Lett. B432 (1998) 139.

\bibitem{PS2}
P. Sikivie, Phys. Rev. D60 (1999) 063501.

\bibitem{Trem}
S. Tremaine, Mon. Not. R. Astron. Soc. 307 (1999) 877.

\bibitem{sim}
J.F. Navarro, C.S. Frenk and S.D.M. White, Ap. J. 462 (1996) 563;
B. Moore et al., Ap. J. Lett. 499 (1998) L5.

\bibitem{2body}
F.S. Labini, T. Baertschiger and M. Joyce, astro-ph/0207029, to appear
in Europhysics Letters; J. Diemand, B. Moore, J. Stadel and 
S. Kazantzidis, MNRAS 348 (2004) 977.

\bibitem{stiff}
D. Stiff and L.M. Widrow, Phys. Rev. Lett. 90 (2003) 211301;

\bibitem{bux} P. Sikivie, astro-ph/9810286, published in the Proceedings
of the Second International Workshop on {\it The Identification of Dark
Matter}, edited by N. Spooner and V. Kudryavtsev, World Scientific 1999,
p.68, astro-ph/9810286.

\bibitem{mw}
P. Sikivie, Phys. Lett. B 567 (2003) 1; L. Duffy and P. Sikivie, 
in preparation.

\bibitem{kin}
W. Kinney and P. Sikivie, Phys. Rev. D61 (2000) 087305.

\bibitem{SW}
P.~Sikivie and S.~Wick, Phys.~Rev.~D66, 023504 (2002).

\bibitem{stream}
D. Stiff, L.M. Widrow and J. Frieman, Phys. Rev. D64 (2001) 083516. 

\bibitem{SagA}
K. Freese, P. Gondolo, H.J. Newberg and M. Lewis, astro-ph/0310334 
and references therein.

\bibitem{ver}
J.D. Vergados, Phys. Rev. D63, (2001) 063511.
%
\bibitem{gre}
A.M. Green, Phys. Rev. D63 (2001) 103003.

\bibitem{gel}
G. Gelmini and P. Gondolo, Phys. Rev. D64 (2001) 023504.

\bibitem{DAMA}
R. Bernabei et al., Riv. N. Cim. 26 n.1 (2003) 1.

\bibitem{cdms}
R. Abusaidi et al., Phys. Rev. Lett. 84 (2000) 5699.

\bibitem{edel}
A. Benoit et al., Phys. Lett. B513 (2001) 15.

\bibitem{zep}
J.C. Barton et al., in the Proceedings of the Fourth International 
Workshop on the Identification of Dark Matter, Sept. 2-6, 2002 in 
York, UK, edited by N.J.C. Spooner and V. Kudryavtsev, World 
Scientific 2003, p 302.

\bibitem{BT}
J. Binney and S. Tremaine, {\it Galactic Dynamics}, Princeton 
University Press, Princeton, NJ, 1987.

\bibitem{eq00}
J. Binney and M. Merrifield, {\it Galactic Astronomy}, Princeton 
University Press, Princeton, NJ, 1998.

\end{references}
\end{document}